\documentclass[twocolumn]{aastex61}

\makeatletter
\newcommand\footnoteref[1]{\protected@xdef\@thefnmark{\ref{#1}}\@footnotemark}
\makeatother

\usepackage{multirow,amsfonts,amsmath,graphicx,natbib,apjfonts,subfigure,color,gensymb,longtable}

\submitjournal{ApJ}

\def \sn{SN\,2016coi}

\def\Msun{M$_{\odot}$}

\def \Ni {$^{56}$Ni}
\def \MNi {$M_{\text{Ni}}$}
\def \Co {$^{56}$Co}
\def \lam {$\lambda$}
\def \Ha {H$\alpha$}

\def \Mej {M_{\rm{ej}}}
\def \Ek {E_k}
\def \Mzams {M$_{\rm{ZAMS}}$}

\DeclareRobustCommand{\ion}[2]{\relax\ifmmode\ifx\testbx\f@series{\mathbf{#1\,\mathsc{#2}}}\else{\mathrm{#1\,\mathsc{#2}}}\fi\else\textup{#1\,{\mdseries\textsc{#2}}}\fi}

\begin{document}

\title{SN\,2016\lowercase{coi} (ASASSN-16\lowercase{fp}): an energetic H-stripped core-collapse supernova from a massive stellar progenitor with large mass loss}

\correspondingauthor{Giacomo Terreran}
\email{giacomo.terreran@northwestern.edu}

\author[0000-0003-0794-5982]{G. Terreran}
\affiliation{Center for Interdisciplinary Exploration and Research in Astrophysics (CIERA) and Department of Physics and Astronomy, Northwestern University, Evanston, IL 60208}

\author[0000-0003-4768-7586]{R. Margutti}
\affiliation{Center for Interdisciplinary Exploration and Research in Astrophysics (CIERA) and Department of Physics and Astronomy, Northwestern University, Evanston, IL 60208}

\author[0000-0001-7485-3020]{D. Bersier}
\affiliation{Astrophysics Research Institute, Liverpool John Moores University, 146 Brownlow Hill, Liverpool L3 5RF, UK}

\author{J. Brimacombe}
\affiliation{Coral Towers Observatory, Cairns, QLD 4870, Australia}

\author{D. Caprioli}
\affiliation{Department of Astrophysical Sciences, Princeton University, 4 Ivy Ln., Princeton NJ 08544, USA}

\author{P. Challis}
\affiliation{Harvard-Smithsonian Center for Astrophysics, 60 Garden Street, Cambridge, Massachusetts, 02138, USA}

\author{R. Chornock}
\affiliation{Astrophysical Institute, Department of Physics and Astronomy, 251B Clippinger Lab, Ohio University, Athens, OH 45701, USA}

\author{D.~L. Coppejans}
\affiliation{Center for Interdisciplinary Exploration and Research in Astrophysics (CIERA) and Department of Physics and Astronomy, Northwestern University, Evanston, IL 60208}
 
\author{Subo Dong}
\affiliation{Kavli Institute for Astronomy and Astrophysics, Peking University, Yi He Yuan Road 5, Hai Dian District, Beijing 100871, China}

\author{C. Guidorzi}
\affiliation{Department of Physics and Earth Science, University of Ferrara, via Saragat 1, I-44122, Ferrara, Italy}

\author{K. Hurley}
\affiliation{University of California, Berkeley, Space Sciences Laboratory, 7 Gauss Way, Berkeley, CA 94720-7450, USA}

\author{R. Kirshner}
\affiliation{Harvard-Smithsonian Center for Astrophysics, 60 Garden Street, Cambridge, Massachusetts, 02138, USA}

\author{G. Migliori}
\affiliation{Dipartimento di Fisica e Astronomia, Alma Mater Studiorum, Universit\`a degli Studi di Bologna, Via Gobetti 93/2, I-40129 Bologna, Italy}
\affiliation{INAF Istituto di Radioastronomia, via Gobetti 101, I-40129 Bologna, Italy}

\author{D. Milisavljevic}
\affiliation{Department of Physics and Astronomy, Purdue University, 525 Northwestern Avenue, West Lafayette, IN 47906, USA}

\author{D.~M. Palmer}
\affiliation{Los Alamos National Laboratory, Los Alamos, NM, 87545, USA}

\author{J.~L. Prieto}
\affiliation{N\'ucleo de Astronom\'ia de la Facultad de Ingenier\'ia, Universidad Diego Portales, Av. Ej\'ercito 441, Santiago, Chile}
\affiliation{Millennium Institute of Astrophysics, Santiago, Chile}

\author{L. Tomasella}
\affiliation{INAF$-$Osservatorio Astronomico di Padova, vicolo dell'Osservatorio 5, Padova I-35122, Italy}

\author{P. Marchant}
\affiliation{Center for Interdisciplinary Exploration and Research in Astrophysics (CIERA) and Department of Physics and Astronomy, Northwestern University, Evanston, IL 60208}

\author{A. Pastorello}
\affiliation{INAF$-$Osservatorio Astronomico di Padova, vicolo dell'Osservatorio 5, Padova I-35122, Italy}

\author{B.~J. Shappee}
\affiliation{Institute for Astronomy, University of Hawai'i, 2680 Woodlawn Drive, Honolulu, HI 96822, USA}

\author{K.~Z. Stanek}
\affiliation{Department of Astronomy, The Ohio State University, 140 West 18th Avenue, Columbus, OH 43210, USA}
\affiliation{Center for Cosmology and Astroparticle Physics, The Ohio State University, 191 W. Woodruff Avenue, Columbus, OH 43210, USA}

\author{M.~D. Stritzinger}
\affiliation{Department of Physics and Astronomy, Aarhus University, Ny Munkegade 120, 8000 Aarhus C, Denmark}

\author{S. Benetti}
\affiliation{INAF$-$Osservatorio Astronomico di Padova, vicolo dell'Osservatorio 5, Padova I-35122, Italy}

\author{Ping Chen}
\affiliation{Kavli Institute for Astronomy and Astrophysics, Peking University, Yi He Yuan Road 5, Hai Dian District, Beijing 100871, China}

\author{L. DeMarchi}
\affiliation{Center for Interdisciplinary Exploration and Research in Astrophysics (CIERA) and Department of Physics and Astronomy, Northwestern University, Evanston, IL 60208}

\author{N. Elias-Rosa}
\affiliation{Institute of Space Sciences (ICE, CSIC), Campus UAB, Carrer de Can Magrans s/n, 08193 Barcelona, Spain}
\affiliation{Institut d’Estudis Espacials de Catalunya (IEEC), c/Gran Capit\'a 2-4, Edif. Nexus 201, 08034 Barcelona, Spain}

\author{C. Gall}
\affiliation{DARK, Niels Bohr Institute, University of Copenhagen, Lyngbyvej 2, 2100 Copenhagen, Denmark}

\author{J. Harmanen}
\affiliation{Tuorla Observatory, Department of Physics and Astronomy, FI-20014 University of Turku, Finland}

\author{S. Mattila}
\affiliation{Tuorla Observatory, Department of Physics and Astronomy, FI-20014 University of Turku, Finland}

\begin{abstract}
We present comprehensive observations and analysis of the energetic H-stripped SN\,2016coi (a.k.a. ASASSN-16fp), spanning the $\gamma$-ray through optical and radio wavelengths, acquired within the first hours to $\sim$420 days post explosion. Our observational campaign confirms the identification of He in the SN ejecta, which we interpret to be caused by a larger mixing of Ni into the outer ejecta layers. From the modeling of the broad bolometric light curve we derive a large ejecta mass to kinetic energy ratio ($M_{\rm{ej}}\sim 4-7\,\rm{M_{\odot}}$, $E_{\rm{k}}\sim 7-8\times 10^{51}\,\rm{erg}$). The small [\ion{Ca}{ii}] \lam\lam7291,7324 to [\ion{O}{i}] \lam\lam6300,6364 ratio ($\sim$0.2) observed in our late-time optical spectra is suggestive of a large progenitor core mass at the time of collapse. We find that SN\,2016coi is a luminous source of X-rays ($L_{X}>10^{39}\,\rm{erg\,s^{-1}}$ in the first $\sim100$ days post explosion) and radio emission ($L_{8.5\,GHz}\sim7\times 10^{27}\,\rm{erg\,s^{-1}Hz^{-1}}$ at peak).
These values are in line with those of relativistic SNe (2009bb, 2012ap). However, for SN\,2016coi we infer substantial pre-explosion progenitor mass-loss with rate $\dot M \sim (1-2)\times 10^{-4}\,\rm{M_{\odot}yr^{-1}}$ and a sub-relativistic shock velocity $v_{sh}\sim0.15c$, in stark contrast with relativistic SNe and similar to normal SNe. Finally, we find no evidence for a SN-associated shock breakout $\gamma$-ray pulse with energy $E_{\gamma}>2\times 10^{46}\,\rm{erg}$. While we cannot exclude the presence of a companion in a binary system, taken together, our findings 
are consistent with a massive single star progenitor
that experienced large mass loss in the years leading up to core-collapse, but was unable to achieve complete stripping of its outer layers before explosion. 

\end{abstract}
\section{Introduction} \label{sec:intro}
\begin{figure*}
\centering
\includegraphics[width=\textwidth]{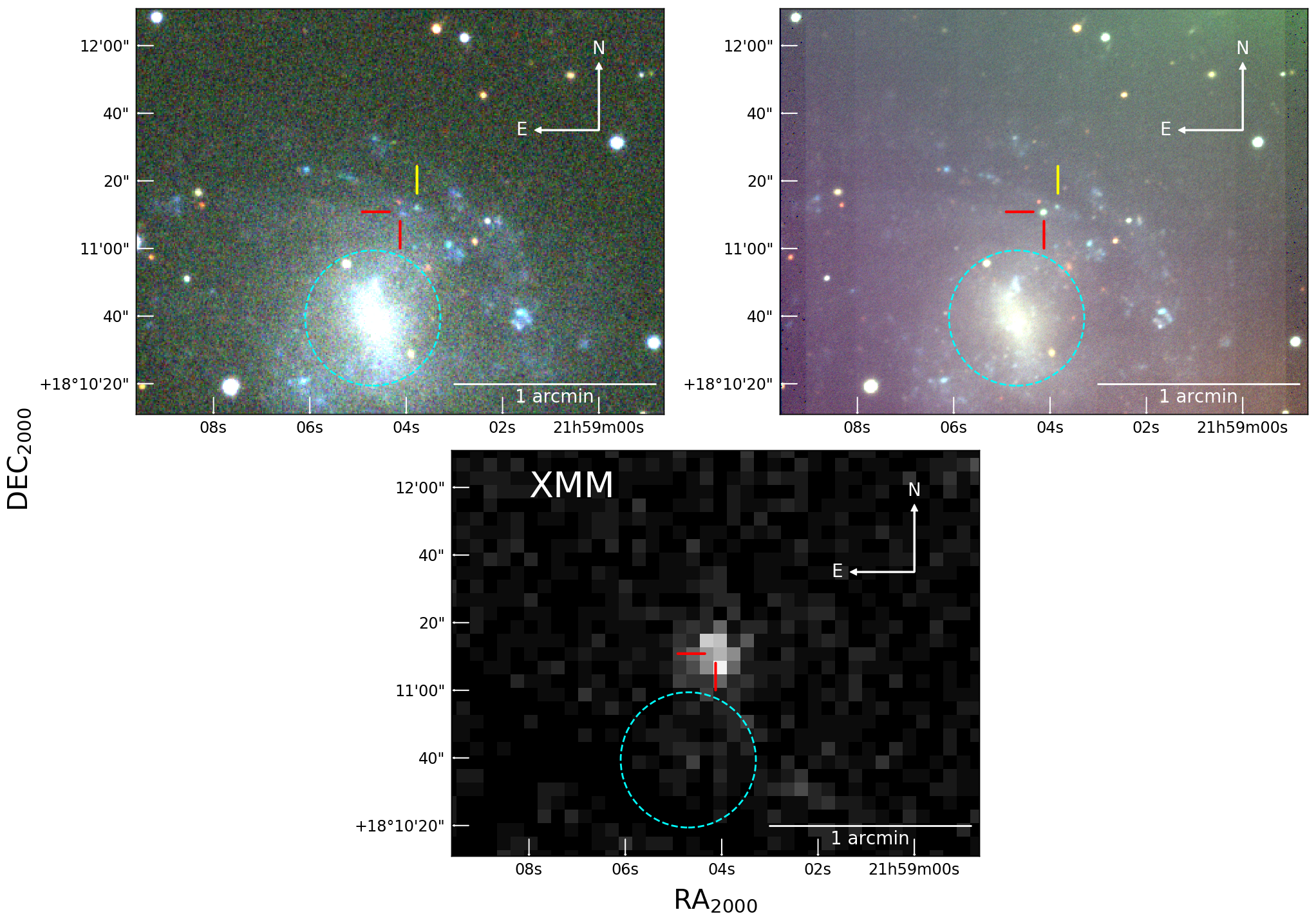}
\caption{Optical (upper panels) and X-ray (lower panel) images of \sn{} and its surroundings. \emph{Upper panels, left}: SDSS pre-explosion false-color image of the host galaxy UGC 11868 of \sn{} in the \textit{gri} filters. Observations were acquired on \mbox{2009, October 16 UT}. \emph{Upper panels, right}: post-explosion false-color image based on \textit{gri} observations acquired with MMTCam on \mbox{2017, June 02 UT} ($\sim1$~yr post explosion). \emph{Lower Panel}: 0.3-10 keV image from XMM-Newton observations at $t\le22$ days. In all panels the location of the SN is marked in red, a yellow mark indicates the location of a nearby \ion{H}{ii} region, while a dashed cyan circle with 20\arcsec\, radius identifies the host-galaxy center.}
\label{Fig:rgb}
\end{figure*}

Hydrogen-stripped core-collapse supernovae (i.e., type Ibc SNe), also called stripped-envelope SNe \citep[SESNe;][]{Clocchiatti1996}, have enjoyed a surge of interest in the last two decades thanks to the association of the most energetic elements of the class with Gamma-Ray Bursts (GRBs). Yet, the stellar progenitors of type Ibc SNe have so far eluded uncontroversial detection in pre-explosion images (\citealt{Gal-Yam2005,Maund2005,Elias-Rosa2013,Eldridge2013}). Relevant in this respect is the discovery of a progenitor in pre-explosion images of the type Ib SN iPTF13bvn, interpreted to be a single Wolf-Rayet (WR) star with a mass at zero-age main-sequence (ZAMS) \Mzams$\sim33$~\Msun{} \citep{Cao2013,Groh2013}. This result was later disputed by \cite{Bersten14}. More recently, \cite{VanDyk2018}, \cite{Kilpatrick2018} and \cite{Xiang2019} identified a source in archival \textit{Hubble Space Telescope} (HST) images covering the location of the type Ic SN\,2017ein, with properties
compatible with a WR star of \Mzams$\sim$$55$~\Msun{} (although the presence of a companion star could not be ruled out).

The stripping of the hydrogen and helium envelope in massive stars mainly occurs through two channels: (i) line-driven winds, which dominate the mass-loss yield in single star evolution; or (ii) interaction with a companion star in a binary system. In the former scenario, the progenitor is expected to be an isolated, massive WR star \citep[\Mzams$\gtrsim20$\Msun;][]{Hamann2006}, consistent with the inferences by \cite{Cao2013}, \cite{Groh2013}, \cite{VanDyk2018} and \cite{Kilpatrick2018} with typical mass-loss rate $\dot M\sim 10^{-5}\,\rm{M_{\sun}\,yr^{-1}}$ \citep{Maeder1981,Woosley1995,Begelman1986}. In the binary progenitor scenario, instead, the primary exploding star is expected to be a helium star (or a C+O star in case of type Ic SNe) with lower-mass \Mzams$\gtrsim12$~\Msun{} \citep{Podsiadlowski1992,Yoon2010a,Eldridge2013,Dessart2015}. The lower mass of the progenitor stars in the binary progenitor scenario naturally accounts for the discrepancy between the large inferred rate of SESNe compared to massive WR stars \citep{Georgy2009,Smith2011,Eldridge2013,Smith2014}, and for the low ejecta masses inferred from the modeling of the bolometric light-curves of type Ibc SNe \citep[$M_{\rm{ej}}\lesssim 3$~\Msun; e.g.,][]{Ensman1988,Drout2011,Dessart2012,Bersten2014,Eldridge2015,Lyman2016,Taddia2018}. In reality, both scenarios are likely contributing in different amounts to the observed population of SESNe.

\begin{table*}
 \centering
\caption{Summary of assumed and inferred parameters from this paper and previous publications.}
\begin{tabular}{lcccc}
\hline
&\cite{Yamanaka17}&\cite{Kumar18}&\cite{Prentice17}&This Work\\
\hline
Distance (modulus $\mu$)&17.2~Mpc (31.18~mag)&18.1~Mpc (31.29~mag)& 15.8~Mpc (31~mag)&$18.1$~Mpc (31.29~mag)\\
Color Excess $E(B-V)_{\rm{tot}}$&0.075~mag&0.074~mag&0.205~mag&0.075~mag\\
Explosion Date &MJD 57532.5& MJD 57533.9&MJD 57533.5&MJD 57531.9\\
Nickel Mass \MNi & 0.15~\Msun & 0.10~\Msun & 0.14~\Msun& 0.15~\Msun\\
Ejecta Mass $\Mej$ & 10~\Msun & $4.5$~\Msun &$2.4-4$~\Msun&$4-7$~\Msun\\
Kinetic Energy $\Ek$ & $3-5\times10^{52}$~erg&$6.9\times10^{51}$~erg&$4.5-7\times 10^{51}$~erg&$7-8\times 10^{51}$~erg\\
He Velocity&$\sim18000$~km s$^{-1}$&$\sim20000$~km s$^{-1}$&$\sim22000$~km s$^{-1}$&$\sim22000$~km s$^{-1}$\\
\hline
\end{tabular}
\label{tab:other_works}
\end{table*}

Here we present the results from an extensive multi-wavelength campaign of the H-poor \sn{} (a.k.a. ASASSN-16fp) from $\gamma$-rays to radio wavelengths, from a few hrs to $\sim 420$ days post explosion. From our comprehensive analysis we infer that \sn{} originated from a compact massive progenitor with large mass loss before explosion, potentially consistent with a single WR progenitor star. 
\sn{} was discovered on 2016 May 27.55 UT \citep[MJD 57535.55;]{Holoien2016} by the All Sky Automated Survey for SuperNovae\footnote{\url{http://www.astronomy.ohio- state.edu/~assassin/index.shtml}} \citep[ASAS-SN;][]{Shappee2014} in the irregular galaxy UGC 11868 (Fig. \ref{Fig:rgb}).
\sn{} was initially classified by the NOT Unbiased Transient Survey \citep[NUTS;][]{Mattila2016} as a type Ic-BL SN similar to those that accompany GRBs \citep{Elias-Rosa2016}, although it was soon realized that traces of He might have been present at early times \citep{Yamanaka2016}.
The optical/UV properties of \sn{} have been studied by \cite{Yamanaka17}, \cite{Prentice17} and \cite{Kumar18}. These authors conclude that \sn{} is an energetic SN with large ejecta mass and spectroscopic similarities to type Ic-BL SNe. In terms of SN classification, \sn{} is intermediate between type Ib and Ic SNe. Unlike type Ib SNe, where He lines become more prominent with time \citep[e.g.,][]{Gal-Yam2017}, the He features of \sn{} disappear after maximum light.

This paper is organized as follows. We first describe our UV, optical and NIR photometry data analysis and derive the explosion properties through modeling of the SN bolometric emission in \S \ref{sec:phot}. Our spectroscopic campaign and inferences on the spectral properties of \sn{} are described in \S \ref{Sec:Spec}. In \S \ref{sec:Radio} we present radio observations of \sn{}, along with the modeling of the blast-wave synchrotron emission, while \S \ref{sec:XRT} is dedicated to the analysis of the luminous X-ray emission of \sn{} and the constraints on the progenitor mass-loss history. We describe our search for a shock breakout signal in the $\gamma$-rays in \S \ref{sec:SBO}. We discuss our findings in the context of properties of potential stellar progenitors in \S\ref{sec:disc} and draw our conclusions in \S\ref{sec:concl}.

In this paper we follow \cite{Kumar18} and adopt $z\simeq0.00365$, which corrected for Virgo infall corresponds to a distance of $18.1\pm1.3$~Mpc ($H_{0} = 73$\,km\,s$^{-1}$\,Mpc$^{-1}$, $\Omega_{\rm M} = 0.27$, $\Omega_{\rm \lambda} = 0.73$), equivalent to a distance modulus of $\mu =31.29\pm 0.15$~mag \citep{Mould2000}. We further adopt a total color excess in the direction of \sn{} $E(B-V)_{\rm{tot}}=0.075$ mag \citep{Schlafly11} as in \cite{Yamanaka17} and \cite{Kumar18}. Unless otherwise stated, time is referred to the inferred time of first light (\S\ref{sec:phot}), which is UT May 23.9 2016 (MJD 57531.9; see \S \ref{sec:Lbol}). The presence of a ``dark phase'' with a duration of a few hours to a few days (e.g., \citealt{Piro13}) has no impact on our major conclusions. Therefore we use the term ``from explosion'' and ``from first light'' interchangeably. A summary of our adopted and inferred parameters is provided in Table \ref{tab:other_works}. Uncertainties are listed at the $1\,\sigma$ confidence level (c.\,l.), and upper limits are provided at the $3\,\sigma$ c.l. unless otherwise noted.

\section{UV, Optical and NIR photometry} \label{sec:phot}
\subsection{Data Analysis} 
\label{subsec:dataUVoptical}

\begin{figure*}[t]
\vskip -0.0 true cm
\centering
\includegraphics[width=\textwidth]{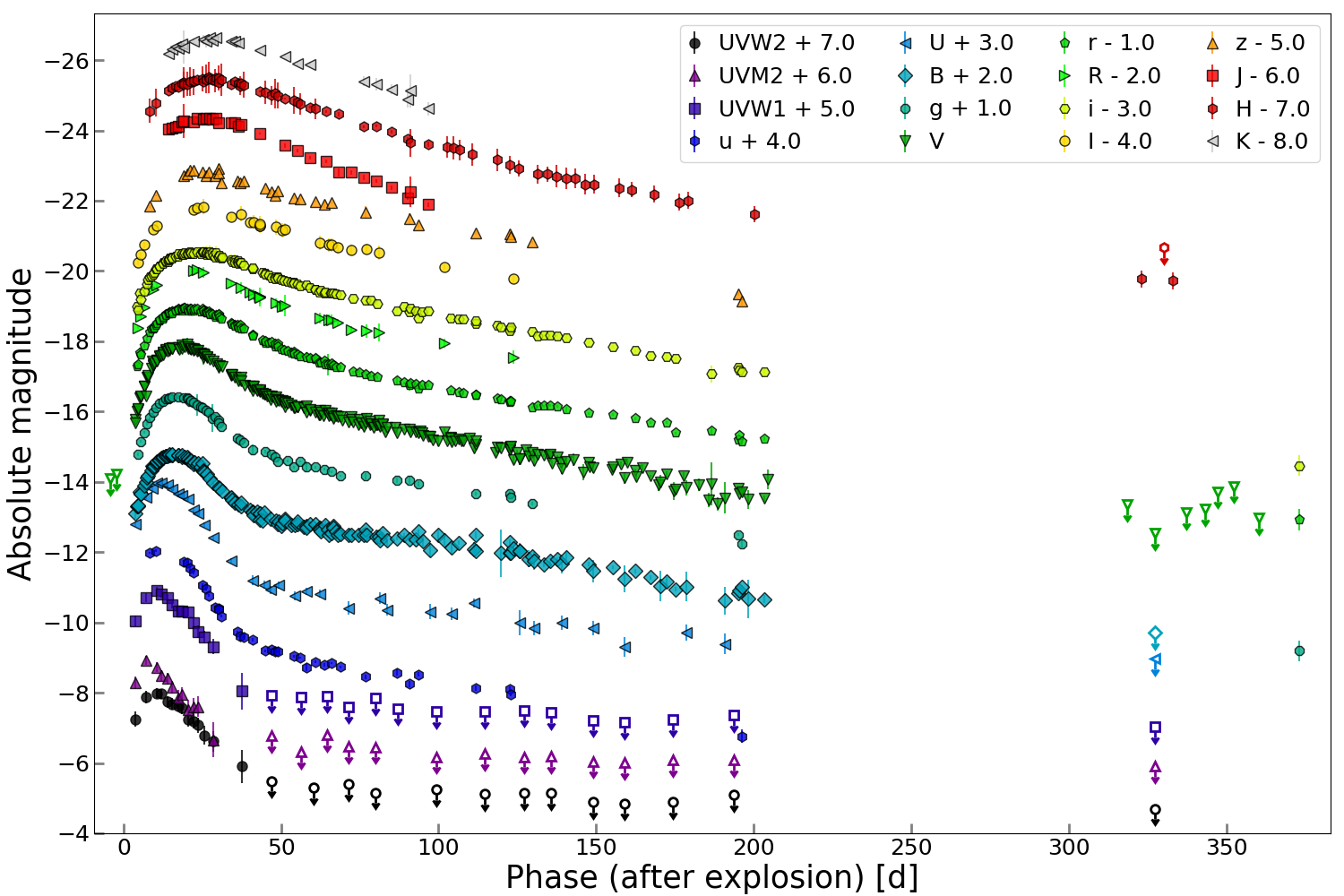}
\caption{UV, optical and NIR emission from \sn{} in the first $\sim$400 days of its evolution. We show extinction corrected absolute magnitudes. 
Upper limits are marked by empty symbols.}
\label{Fig:snlcall}
\end{figure*}

Our photometric data have been obtained from several different telescopes and instruments, which are listed in Table \ref{tab: instr}.
UV data have been acquired with the Ultraviolet Optical Telescope \citep[UVOT;][]{Roming2005}, on the Neil Gehrels \emph{Swift} Observatory \citep{Gehrels2004}.
We measured the SN instrumental magnitudes by performing aperture photometry with the \texttt{uvotsource} task within the \textsc{HEAsoft} v6.22,
and following the guidelines in \cite{Brown2009}. An aperture of 3\arcsec{} was used. We estimated the level of contamination from the host galaxy flux 
using late-time observations acquired at $t\sim322$~d after first light, when the SN contribution is negligible.
We then subtracted the measured count-rate at the location of the SN from the count rates in the SN images following the prescriptions in \cite{Brown2014}.

Images acquired with the Liverpool Telescope have been processed with a custom-made pipeline, while we use standard overscan, bias and flatfielding within \textsc{iraf}\footnote{\textsc{IRAF} is distributed by the National Optical Astronomy Observatory, which is operated by the Association of Universities for Research in Astronomy (AURA) under a cooperative agreement with the National Science Foundation. http://iraf.noao.edu/} for the remaining optical photometry.
NOTCam NIR images were reduced with a modified version of the external \textsc{IRAF} package \textsc{IRAF} (v. 2.5)\footnote{\url{http://www.not.iac.es/instruments/notcam/guide/observe.html\#reductions}}.
The remaining NIR data reduction has been performed through standard flat-field correction, sky background subtraction and stacking of the individual exposures for an improved signal-to-noise ratio.
The photometry has been extracted using the \textsc{SNOoPY}\footnote{Cappellaro, E. (2014). \textsc{SNOoPY}: a package for SN photometry, \url{http://sngroup.oapd.inaf.it/snoopy.html}} package. We performed point-spread-function (PSF) photometry with \textsc{DAOPHOT} \citep{Stetson1987}.
For non-detections we calculated upper limits corresponding to a S/N of 3. Zero points and color terms for each night have been estimated based on the magnitudes of field stars in the Sloan Digital Sky Survey\footnote{\url{http://www.sdss.org}} \citep[SDSS;][]{York2000} catalog (DR9). 
We converted the SDSS \textit{ugriz} magnitudes to Johnson/Cousins \textit{UBVRI} filters following \cite{Chonis2008}. For NIR images, we used the Two Micron All Sky Survey (2MASS) catalog\footnote{\url{http://www.ipac.caltech.edu/2mass/}} \citep{Skrutskie2006}. 
We quantified the uncertainty on the instrumental magnitude injecting artificial stars
\citep[e.g.,][]{Hu2011}. The resulting uncertainty was then added in quadrature to the fit uncertainties returned by \textsc{DAOPHOT} and the uncertainties from the photometric calibration to obtain the total uncertainty on the photometry. Our final values are reported in Tables \ref{fig: ugriz}$-$\ref{fig: SDA} and shown in Fig. \ref{Fig:snlcall}.

Our UV-to-NIR campaign densely samples the evolution of \sn{} in its first $\sim400$ days post explosion, with more than 1100 observations distributed over 166 nights (the gap around 200-300 days corresponds to when \sn{} was behind to the Sun). As Fig. \ref{Fig:snlcall} shows, \sn{} rises to peak considerably faster in the bluer bands. The UV filters also show the fastest decline post peak, before relaxing on a significantly slower decay at $t\gtrsim40$. This sharp change of decay rate is not present in the redder bands, which instead show a roughly constant decay rate after peak. The late-time $V$-band decays as 1.7~mag 100d$^{-1}$, faster than expected from the radioactive decay of \Co{}, suggesting leakage of $\gamma$-rays. Our last detections of \sn{} at $\sim373$~d post explosion are consistent with the temporal decay inferred from earlier observations at $50\rm{d}\lesssim t\lesssim300 {d}$ (Fig. \ref{Fig:snlcall}). Finally, by using a low-order polynomial fit we measure the time of maximum light in the V band V$_{max}=18.34\pm0.16$~d after discovery, corresponding to MJD 57550.24 (2016 June 11.24 UT). The time of peak in other bands is reported in Table \ref{tab: max}.

\subsection{Bolometric Luminosity and Explosion Parameters} \label{sec:Lbol}
\begin{figure}
 \centering
\includegraphics[width=\columnwidth]{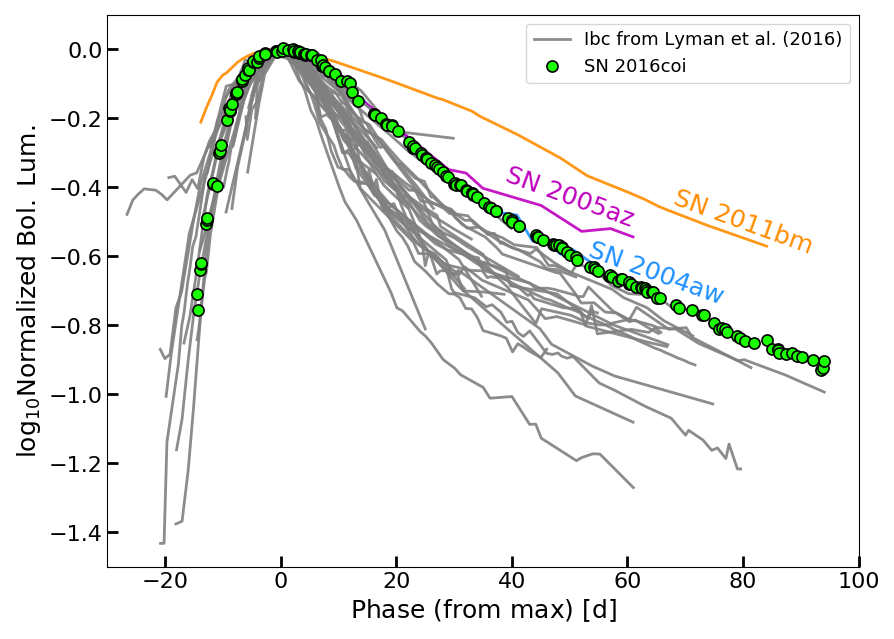}
 \caption{Comparison of the \textit{uvoir} bolometric light curve of \sn{} with the sample of SESNe from \cite{Lyman2016}. The light curves have been normalized to maximum light. \sn{} is among the objects with the broadest light curve, suggesting a larger than average diffusion time. The broad light curves of SNe 2004aw \citep[which lies exactly below \sn{};][]{Taubenberger2006}, 2005az \citep{Drout2011} and 2011bm \citep{Valenti2012} are also marked.}
\label{Fig:LCcomparison}
\end{figure}

Our extensive photometric coverage allows us to reconstruct the bolometric emission from \sn{} from the UV to the NIR from a few days to $\sim200$ days after explosion. As a comparison, the bolometric light-curve from \cite{Prentice17} has similar temporal coverage but does not include the NIR and UV contributions, while \cite{Kumar18} and \cite{Yamanaka17} include either the UV emission or the NIR emission until $\delta t\le 60$ days post explosion, respectively. We build the bolometric luminosity curve of \sn{} starting from extinction-corrected flux densities, and we interpolate the flux densities in each filter to estimate the SN emission at any given time of interest. In case of incomplete UV-to-NIR photometric coverage we assume constant color from the previous closest epochs. Finally, we integrate the resulting spectral energy distributions (SEDs) from the UV to the NIR with the trapezoidal rule to obtain the bolometric light-curve shown in Fig. \ref{Fig:LCcomparison}.

We compare the bolometric light curve of \sn{} with a sample of well-observed H-stripped core-collapse SNe from \cite{Lyman2016} in Fig. \ref{Fig:LCcomparison}. \cite{Lyman2016} used the parameter $\Delta_{15}$ as an estimator of the broadness of the light curve, defined as the the difference in magnitude between the luminosity at peak at the luminosity 15~d after that. The smaller the $\Delta_{15}$, the ``slower'' the event, i.e., the broader the light curve. The SN with the broadest light curve in the sample of \cite{Lyman2016} is the type-Ic SN\,2011bm, which has $\Delta_{15}=$0.2~mag. Other slow events are the type-Ic SNe 2004aw, 2005az (with $\Delta_{15}$=0.41 and $\Delta_{15}$=0.42~mag, respectively) and the type-Ib SNe 1999dn and 2004dk (with $\Delta_{15}$=0.32 and $\Delta_{15}$=0.41~mag, respectively). Figure \ref{Fig:LCcomparison} shows that with $\Delta_{15}$=0.41~mag, \sn{} is among the SNe with the broadest light-curves. \cite{Kumar18} did a similar analysis looking at the $\Delta_{15}$ in the single bands and reached the same conclusion.

The broad light curve indicates a large photon diffusion time scale, and hence a large ejecta mass ($M_{ej}$) to kinetic energy ($E_{k}$) ratio. Assuming standard energetics, this translates to a considerably large ejecta mass, in agreement with previous findings by \cite{Prentice17} and \cite{Kumar18}. Interestingly, \sn{} shows a very slow post-peak decline with a standard time to peak $t_{rise}<20$ days (Fig. \ref{Fig:LCcomparison}). This phenomenology might result from mixing of $^{56}$Ni in the outer stellar ejecta, as opposed of having all the \Ni{} located at the centre of the explosion. We quantify these statements below.

We model the bolometric light-curve of \sn{} adopting the formalism by \cite{Arnett82} modified following \cite{Valenti08} and \cite{Wheeler15}. We adopt a mean opacity $\kappa_{opt}=0.07\,\rm{cm^{2}g^{-1}}$ and break the model degeneracy using a photospheric velocity $v_{phot}\sim16000\,\rm{km\,s^{-1}}$ around maximum light, as inferred from \ion{Fe}{ii} spectral lines (\S \ref{Sec:Spec}). We find that during the photospheric phase at $t<30$ days the light-curve is well described by a model with kinetic energy $E_{k,phot}\sim7\times 10^{51}\,\rm{erg}$, \Ni{} mass $M_{Ni,phot}=0.13\,\rm{M_{\sun}}$ and ejecta mass $M_{ej,phot}\sim4\,\rm{M_{\sun}}$, consistent with the findings of \cite{Kumar18}. However, this model significantly underestimates the bolometric emission during the nebular phase. This is a common outcome of the modeling of energetic type-Ic SN light-curves, which motivated \cite{Maeda03} and \cite{Valenti08} to consider a two-component model. In two-component models the ``outer component'' dominates the early-time emission during the photospheric phase, while the late-time nebular emission receives a significant contribution from a denser inner core (``inner component''). Applying this modeling we find a total ejecta mass $M_{ej}\sim (4-7)\,\rm{M_{\sun}}$, $E_{k}\sim (7-8)\times 10^{51}\,\rm{erg}$ and $M_{Ni}\sim 0.15\,\rm{M_{\sun}}$, with a larger fraction of \emph{Ni} per unit mass in the outer component. This model also allows us to constrain the time of first light to MJD $57531.9\pm 1.5$ days (May 23.9, 2016 UT). 

As a comparison, the spectral modeling by \cite{Prentice17} indicates $M_{ej}=2.4-4\,\rm{M_{\sun}}$, $E_k=(4.5-7)\times 10^{51}\,\rm{erg}$. Scaling the emission of \sn{} to the GRB-associated SN 2006aj and SN\,2008D, \cite{Yamanaka17} find $M_{ej}\sim 10\,\rm{M_{\sun}}$, $E_k=(3-5)\times 10^{52}\,\rm{erg}$ (Table \ref{tab:other_works}). The rough agreement among the results is not surprising given the very different methods used (with different assumptions) and the fact that the modeling of \cite{Prentice17} is limited to optical spectra, and \cite{Yamanaka17} only consider the optical/NIR emission of \sn{} during the early photospheric phase.

\section{Optical Spectroscopy}
\label{Sec:Spec}

\subsection{Data Analysis}
\label{SubSec:dataAnSpec}

\begin{figure*}
\vskip -0.0 true cm
\centering
\includegraphics[height=\dimexpr \textheight - 4\baselineskip\relax]{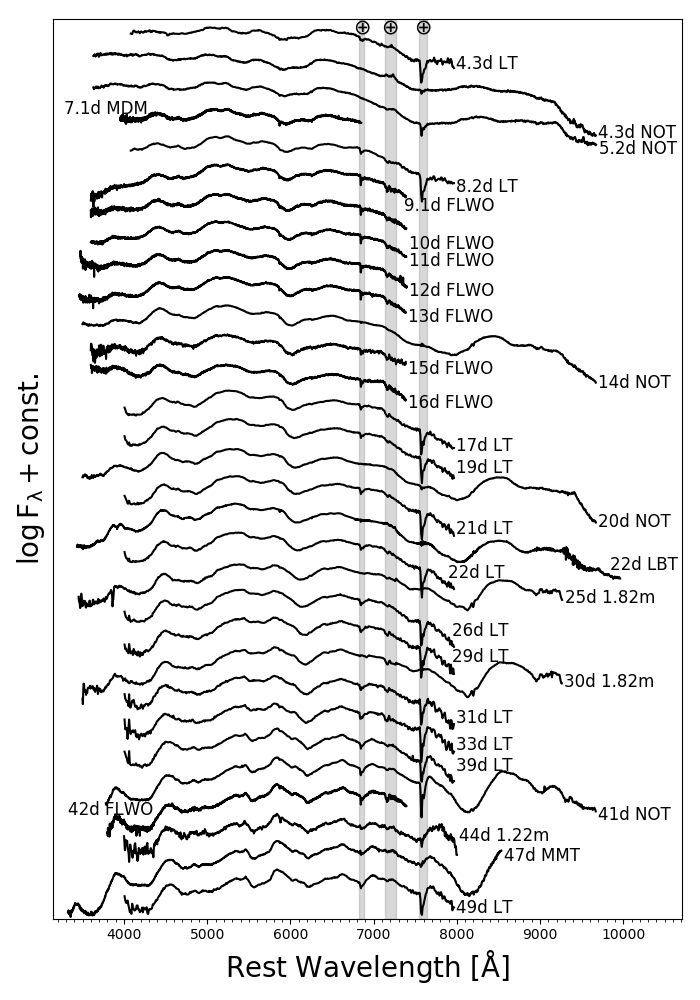}
\caption{Optical spectral evolution of \sn{}. The spectra are presented in the rest-frame ($z=0.003646$) and have been corrected for Galactic extinction along the line of sight. The spectra are shifted vertically for displaying purposes. Spectra are labeled based on the epoch of their acquisition and telescope used. The gray vertical bands mark the positions of the telluric O$_2$ A and B absorption bands.}
\label{Fig:spec}
\end{figure*}

\addtocounter{figure}{-1}

\begin{figure*}
\vskip -0.0 true cm
\centering
\includegraphics[height=\dimexpr \textheight - 4\baselineskip\relax]{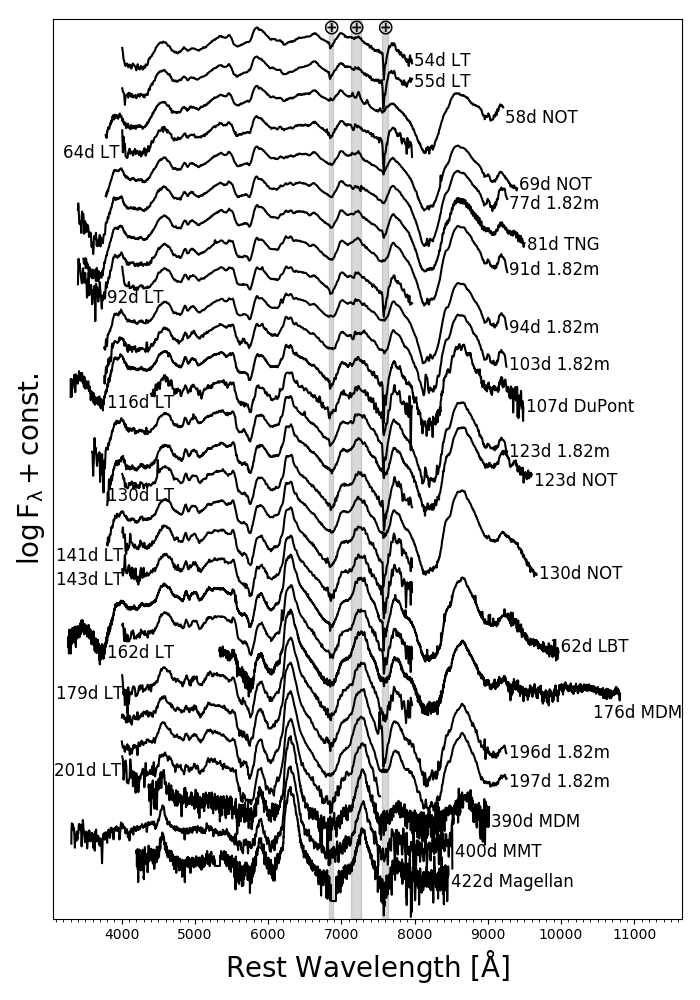}
\caption{\textit{Continued:} Optical spectral evolution of \sn{}. The spectra are presented in the rest-frame ($z=0.003646$) and have been corrected for Galactic extinction along the line of sight. The spectra are shifted vertically for displaying purposes. Spectra are labeled based on the epoch of their acquisition and telescope used. The gray vertical bands mark the positions of the telluric O$_2$ A and B absorption bands.}
\end{figure*}

\begin{figure}[t]
\vskip -0.0 true cm
\centering
\includegraphics[width=\columnwidth]{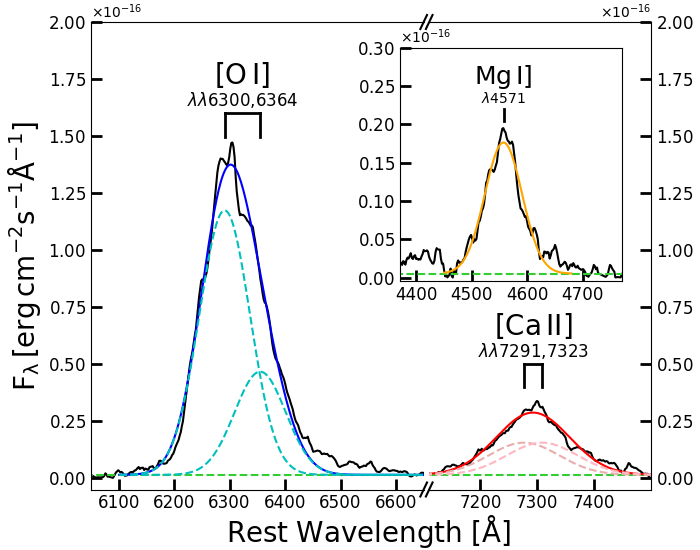}
\caption{A zoomed-in plot of the spectral region of the [\ion{O}{i}] \lam\lam6300,6364, [\ion{Ca}{ii}] \lam\lam7291,7323 and \ion{Mg}{i}] \lam4571 lines of the MMT+BlueChannel nebular spectrum acquired on 2017, June 28 ($\sim400$~d after explosion). Gaussian profiles have been used to reconstruct the doublet components of the emission features. The observed lack of asymmetry of the [\ion{O}{i}] emission feature might result from spherically symmetric ejecta, or possibly from an axisymmetric explosion, viewed at an angle below $50^\circ$.}
\label{fig:neb}
\end{figure}

\begin{figure*}[t]
\vskip -0.0 true cm
\centering
\includegraphics[width=\textwidth]{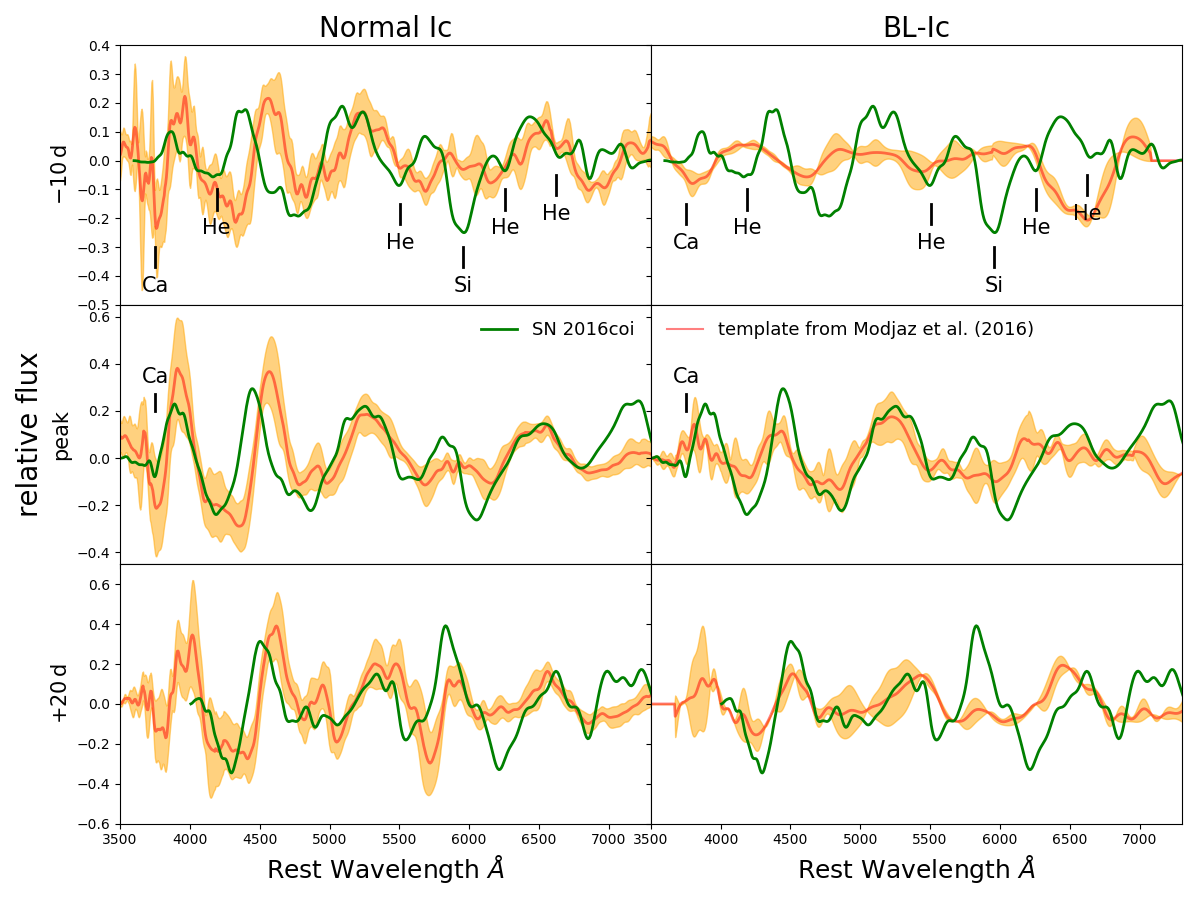}
\caption{Comparison between the spectra of \sn{} with the spectral templates from \cite{Modjaz2016} at $-10$~d from peak, maximum light and $+20$~d after peak (top, middle and bottom panels, respectively). We compare \sn{} to normal type Ic-SNe (left panels) and with Ic-SNe with broad lines (i.e., type BL-Ic; right panels). The orange shaded region represents a $1\,\rm{\sigma}$ standard deviation from the mean. \sn{} shares similarities with both classes, and it can be considerate as an intermediate case that bridges the gap between type Ic and BL-Ic SNe.}
\label{fig:modjaz}
\end{figure*}

We obtained optical spectroscopy of \sn{} from a few days until $t>400$ days post explosion with a variety of instruments on different telescopes (Table \ref{tab: instr2}). The spectroscopic log can be found in Table \ref{fig: spec_log}.
We extracted our time series of optical spectra with \textsc{IRAF} following standard procedures.
Comparison lamps and standard stars acquired during the same night and with the same instrumental setting have been used for the wavelength and flux calibrations, respectively.
When possible, we further removed telluric bands using standard stars.

Our spectroscopic campaign comprises 65 spectra (Fig. \ref{Fig:spec}). The overall evolution of \sn{} is similar to that of type-Ic SNe. At early times $t\lesssim60$~d the blue part of the spectrum at $\lambda\lesssim 5500$~\AA{} is dominated by blends of several \ion{Fe}{ii} multiplets. We identify the main spectral feature at $\sim6000$~\AA{} as \ion{Si}{ii} \lam6355. Before maximum light, we associate the absorption feature around $\sim5500$~\AA{} to \ion{He}{i} \lam5876, with possible contamination by \ion{Na}{i} D. \ion{Na}{i} dominates after maximum light. 
At $\lambda>7000$~\AA{} the spectra of \sn{} show emission associated with \ion{O}{i} \lam\lam7771,7774,7775 and the \ion{Ca}{ii} NIR triplet. Nebular features start to appear $\sim90$~d after explosion, when the forbidden [\ion{O}{i}] \lam\lam6300,6364 and the [\ion{Ca}{ii}] \lam\lam7291,7323 doublets begin to emerge. 

In Fig. \ref{fig:neb} we show a zoomed-in plot of the nebular spectrum acquired with MMT+BlueChannel at $\sim400$~d after explosion. We plot the region of the forbidden [\ion{O}{i}] \lam\lam6300,6364, [\ion{Ca}{ii}] \lam\lam7291,7323 doublets, and semi-forbidden \ion{Mg}{i}] \lam4571 emission line. We use gaussian profiles to model each emission line. For the doublets, we kept the separation between the two components fixed, while allowing for rigid shifts of the overall profile (this scheme will also be followed in \S \ref{subsec:broad LC}). This simple approach allows us to adequately reproduce the emission line profiles (Fig. \ref{fig:neb}). 
We find that the ratio between the oxygen lines fluxes is $\sim2.6$, in reasonable agreement with the theoretical expectation of $\sim3$. However, the doublet is blue-shifted by $\sim10$~\AA{} ($\sim400$~km\,s$^{-1}$). We find similar blue-shifts for the Ca and Mg lines. 
Blue-shifted oxygen line profiles of this kind are not uncommon in type Ibc SN nebular spectra, and several causes have been invoked to explain this observed phenomenology, including dust obscuration, internal scattering, contamination from other lines or residual opacity in the core of the ejecta \citep{Modjaz2008b,Taubenberger2009,Milisavljevic2010}.
We do not observe asymmetric structures in the spectral lines, nor do we detect any sharp decrease in the light curve of \sn{}, therefore we can confidently exclude the presence of dust \citep{Elmhamdi2003,Elmhamdi2004}. As the [\ion{O}{i}] forbidden doublet is fairly isolated, we disfavour contamination from other lines as the origin for the blue-shift. The fact that lines of different species show this behaviour might suggest a geometrical effect. An asymmetric explosion with a bulk of material moving towards the observer could indeed cause the blue-shift. Qualitative inferences on the geometry and distribution of the oxygen-rich ejecta in \sn{} can be drawn from the line profile of the forbidden [\ion{O}{i}] \lam\lam6300,6364 \citep{Modjaz2008b,Taubenberger2009,Milisavljevic2010}. Double-peaked oxygen lines are usually interpreted to be formed in asymmetric explosions viewed at a high angle between the observer point of view and the jet direction \citep{Maeda08,Taubenberger2009}. As shown in Fig. \ref{fig:neb}, the oxygen doublet in \sn{} presents a single, symmetric profile, reproducible with simple Gaussian functions. This result is consistent with spherically symmetric ejecta. As \cite{Maeda08} have shown that an asymmetric profile would not develop for asymmetric explosions viewed from angles below $\sim50^\circ$, an asymmetric explosion cannot be ruled out. However, the asymmetric explosion scenario might actually be supported by the red excess visible in both the oxygen and magnesium line. Indeed, magnesium and oxygen are expected to have similar spatial distribution within the SN ejecta \citep[e.g.,][]{Maeda06,Taubenberger2009}. Such an excess, visible in both features, is unlikely to be caused by line contamination, and is rather the result of ejecta asymmetries common to both line emission regions.

We conclude with a consideration on intrinsic reddening. In our highest resolution spectra acquired on November 2, 2016 UT ($\sim162$ days after first light) with LBT+MODS, we find a weak narrow \ion{Na}{i} D absorption at the redshift of the host galaxy, from which we infer $E(B-V)_{\rm{host}}\sim0.017$ mag \citep{Turatto2003b,Poznanski2012}. However, given the large uncertainties of this method \citep{Phillips2013}, and the lack of evidence for significant $E(B-V)_{\rm{host}}$, in the following we assume $E(B-V)_{\rm{host}}=0$ mag.\footnote{This is in agreement with the assumption by \cite{Yamanaka17} and \cite{Kumar18}. On the other hand, \cite{Prentice17} assumed a host extinction of $E(B-V)_{\rm{host}}$=0.125 mag.} This assumption has no impact on our conclusions. Following \cite{Schlafly11}, the Milky Way color excess in the direction of \sn{} is $E(B-V)_{\rm{MW}}$=0.075~mag, which we use to correct our spectro-photometric data for extinction.

\subsection{SN Classification and presence of Helium in the ejecta}
\label{sec: class}

\sn{} initially showed spectral similarities to type Ic-BL SNe (and in particular to SN\,2006aj, associated with GRB\,060218) but later evolved to resemble a normal type Ic SN (Fig. \ref{Fig:spec}). Indeed, both \cite{Kumar18} and \cite{Prentice17} identified \sn{} as being an intermediate object between the two classes, while \cite{Yamanaka17} classified \sn{} as a BL-Ib SN, because of the presence of helium in the spectra and expansion velocities larger than in normal type Ib SNe. We quantitatively explore the questions of how the ejecta velocity of \sn{} compares to other H-stripped SNe, and the presence of He in its ejecta below.

We compare \sn{} to the spectral templates of normal type Ic SNe and BL-Ic SNe from \cite{Modjaz2016} in Fig. \ref{fig:modjaz} after applying 
the same renormalization procedure. The result is presented for three different epochs: 10~days before maximum light, around maximum light, and 20~days after maximum light. Fig. \ref{fig:modjaz} demonstrates that the typically prominent \ion{Ca}{ii} H+K absorption feature of type Ic SNe spectra is almost absent in \sn{} (Fig. \ref{fig:modjaz}, top and middle panels), in closer similarity to type Ic-BL SN spectra. Notably, \sn{} shows a prominent absorption feature at $\sim6000$~\AA{} that we identify as \ion{Si}{ii} with $v\sim19000$~km s$^{-1}$, which is typically not present with this strength in normal type Ic SNe \citep[e.g.,][]{Parrent2016}.

From our comparison, \sn{} more closely resembles normal type Ic SNe, especially before maximum light.
Compared to type Ic-BL SNe, \sn{} shows more prominent peaks and troughs (upper panels in Fig. \ref{fig:modjaz}), as a result of its lower ejecta velocities before maximum light, which cause less severe blending of the spectral features. Compared to normal type Ic SNe, however, \sn{} shows systematically blue-shifted spectral features. \cite{Modjaz2016} showed that in type Ic SNe (both normal and broad-line) the ``broadness'' of the spectral features correlates with the blue-shift of their minima, as is expected from an expanding atmosphere \citep[e.g.,][]{Dessart2011}. However, with very blue-shifted absorption minima similar to type Ic-BL, but less prominent broadening, \sn{} seems to deviate from this trend.

\cite{Modjaz2016} used the \ion{Fe}{ii} $\lambda$5169 to show this correlation between the blue-shift of the minima and the broadening of the absorption feature. We use the same fitting technique as in \cite{Modjaz2016} to measure the broadening of this same line for \sn{} at maximum light, obtaining $v_{broad}\sim2380$~km~s$^{-1}$. Comparing this value with their Fig. 7, it is possible to see how this is quite low for a BL-Ic, while the velocity inferred from the position of the minimum of the line profile is $v_{min}\sim18050$~km~s$^{-1}$, well within the range of the other BL-Ic of their sample. Another event that had very blue-shifted minima but relatively low broadening was PTF\,12gzk \citep{Ben-Ami12}. These observations were interpreted as resulting from either the departure from spherical symmetry, or from a steep gradient of the density profile of the progenitor envelope. Interestingly, \cite{Ben-Ami12} inferred a massive ejecta of 25-35\Msun{} and a large kinetic energy of $5-10\times 10^{51}$~erg for PTF\,12gzk, which is comparable to \sn{}. 
\sn{} thus shows spectral properties that are intermediate between type Ic-BL SNe (with which \sn{} also shares the large kinetic energy $E_{k}>10^{51}\,\rm{erg}$ but lower velocities before peak) and normal type Ic SNe. These results agree with the findings by \cite{Kumar18} and \cite{Prentice17}.

\begin{figure}
\vskip -0.0 true cm
\centering
\includegraphics[width=\columnwidth]{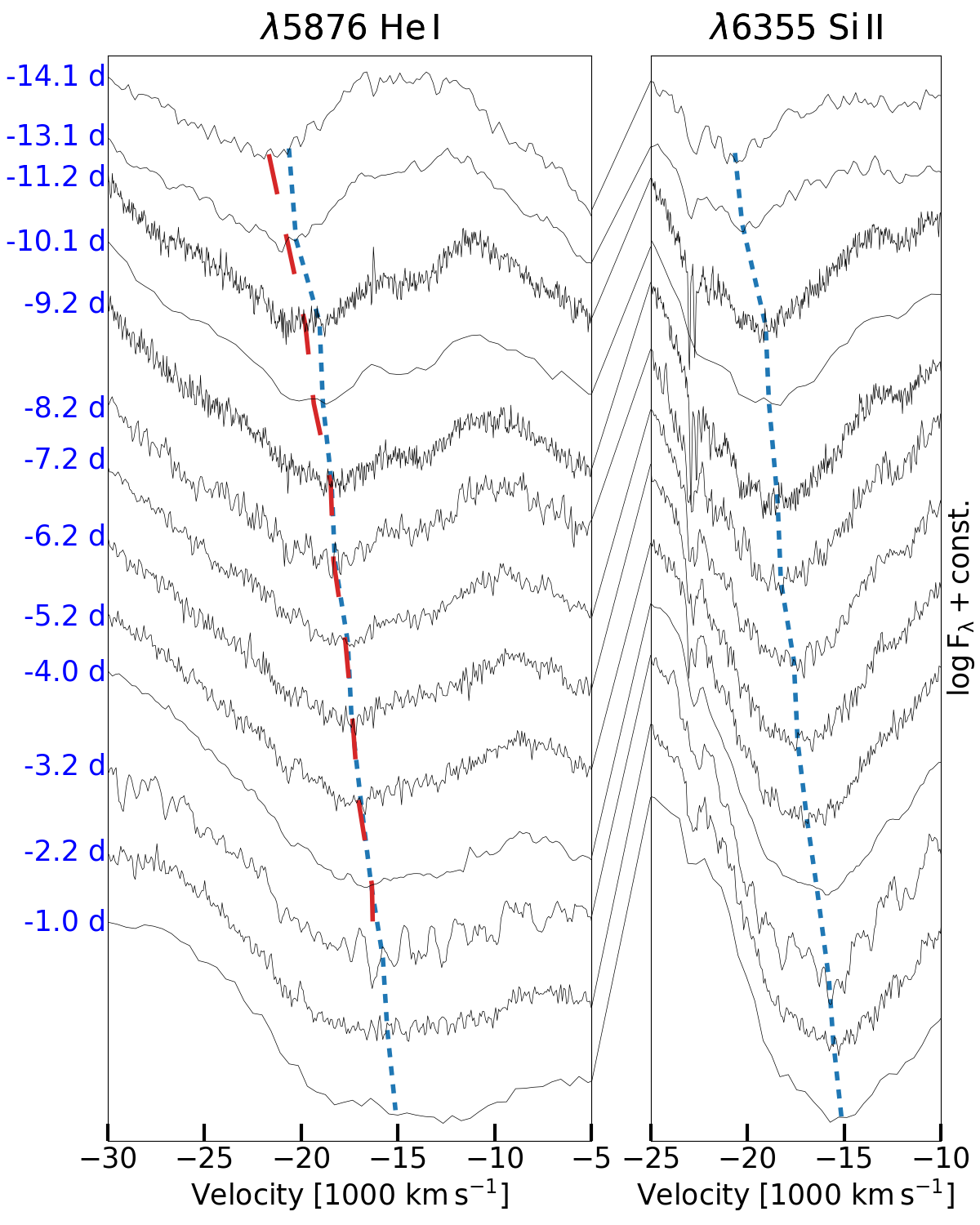}
\caption{\textit{Right panel:} Evolution of the \ion{Si}{ii} \lam6355 line before maximum light in the velocity space. The position of the minimum of the absorption feature is marked with a vertical blue short-dashed line. \textit{Left panel:} Evolution of the 5500~\AA feature, assumed to be the \ion{He}{i} \lam5876 line. The position of the minimum of the absorption feature is marked with a red long-dashed line. The velocity of the absorption minimum of \ion{Si}{ii} \lam6355 is also shown for comparison with a blue short dashed line. The match between the evolution of the two lines suggest a correct interpretation of the 5500~\AA feature as the \ion{He}{i} \lam5876 line. By the time of maximum light, He absorption is no longer apparent in the spectra of \sn{}, and the presence of He becomes hard to quantify due the possible emergence of the \ion{Na}{i} \lam\lam5890,5896 doublet.}
\label{Fig:He_ev}
\end{figure}

We next address the presence of He in the ejecta of \sn{} (in Fig. \ref{fig:modjaz} we marked the position of the \ion{He}{i} \lam4472, \lam5876 \lam6678 and \lam7065 lines). We investigate the velocity evolution of the most prominent spectral features among those associated with \ion{He}{i} at 5876~\AA{} in Fig. \ref{Fig:He_ev}, and use the velocity evolution inferred from \ion{Si}{ii} \lam6355 as a comparison. From Fig. \ref{Fig:He_ev} we find that He and Si show a very similar temporal evolution, with expansion velocities evolving from $v\sim20000\,\rm{km\,s^{-1}}$ at $\sim$2 weeks before maximum light, to $v\sim15000\,\rm{km\,s^{-1}}$ around peak. The identification of He might inspire a connection with type Ib SNe. However, we note that in \sn{} the He features slowly subside (by the time of maximum light He absorption is no longer prominent, Fig. \ref{Fig:He_ev}), while in type Ib SNe He features develop with time \citep[e.g.,][]{Filippenko97,Gal-Yam2017}.
The presence of He in \sn{} has been recognized as a peculiar characteristic of \sn{} by \cite{Yamanaka17}, \cite{Kumar18} and \cite{Prentice17}. \cite{Yamanaka17} concluded the presence of He in \sn{} based on the comparison with a smoothed out and blueshifted version of the type Ib SN 2012au \citep{Takaki2013}, finding a correspondence with the position of the main helium features. They also cross-checked this result with synthetic spectra generated with the code SYN++ \citep{Thomas2011}. \cite{Kumar18} adopted a similar strategy to the one presented in this work, performing a detailed velocity analysis of the single features. With their 1D Monte Carlo spectra synthesis code, \cite{Prentice17} investigated what other elements could be responsible for the absorption at $\sim5500$~\AA{}. They showed that He is indeed the favored interpretation, and that in the absence of He, unphysical amounts of \ion{Al}{ii} and \ion{Na}{i} would be necessary to reproduce the observed spectra.

Similar velocities between Si-rich and He-rich ejecta indicates a clear departure from the expectations of a homologous explosion of a stratified progenitor star where the outer He-rich layers are expected to expand significantly faster than the inner Si-rich layers of ejecta. This finding suggests a higher level of mixing of the ejecta, which might be connected with the capability to excite He (and hence the detection of He in our spectra).

\section{Radio}\label{sec:Radio}
\begin{figure}
\vskip -0.0 true cm
\centering
\includegraphics[width=\columnwidth]{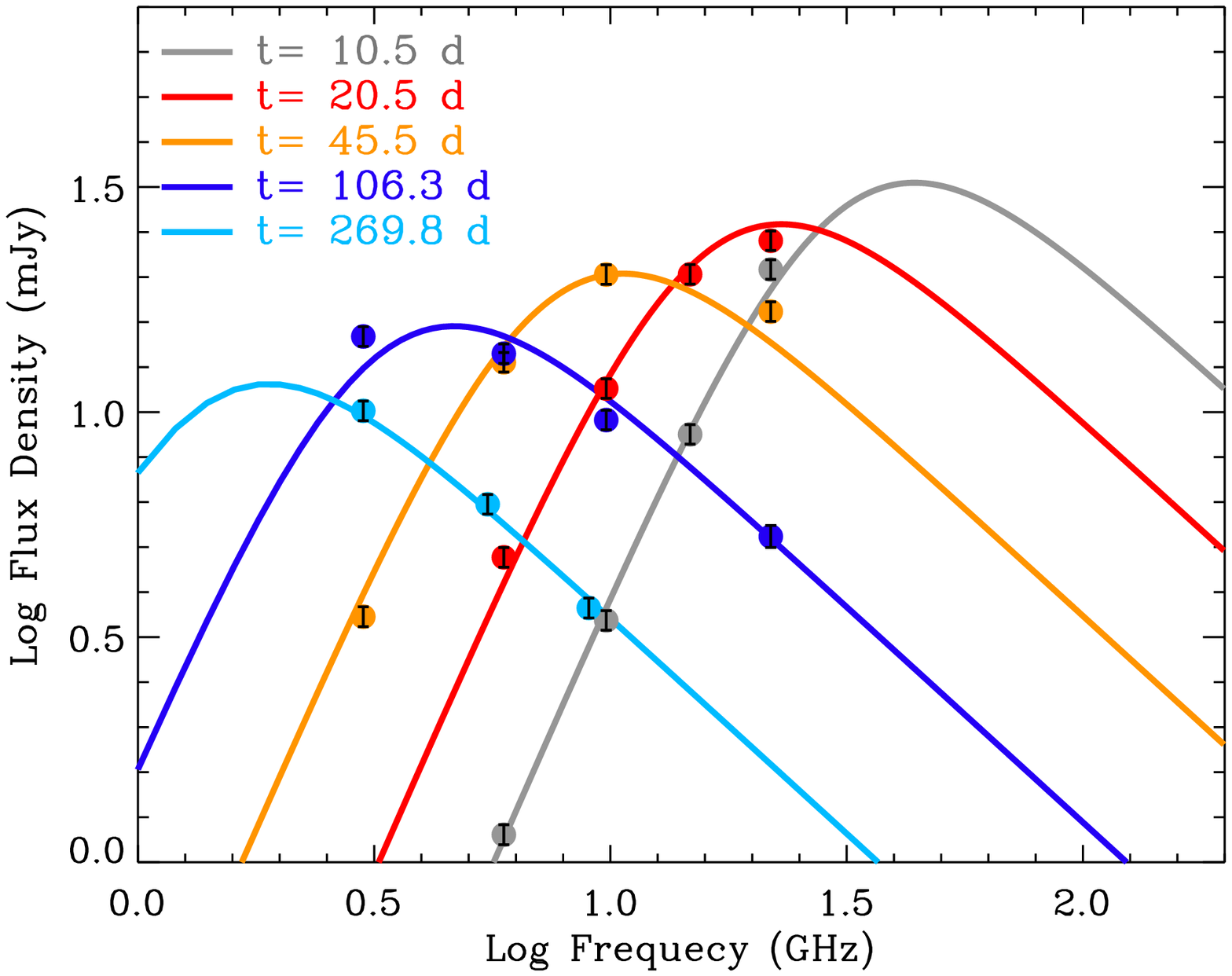}
\caption{Radio SED of \sn{} at 10.5, 20.5, 45.5, 106.3 and 269.8 days after first light (see Table \ref{Tab:radio}). The radio emission from \sn{} is well described by a synchrotron self-absorbed spectrum (SSA) with spectral peak frequency $\nu_{pk}\propto t^{-0.97\pm0.02}$ and peak flux $F_{pk}\propto t^{-0.31 \pm 0.02}$. We find $F_{\nu}\propto \nu^{2.4 \pm 0.1}$ for the optically thick part of the spectrum, consistent with $F_{\nu}\propto \nu^{5/2}$ as expected for SSA. The optically thin part of the spectrum $F_{\nu}\propto \nu^{-(p-1)/2}$ scales as $F_{\nu}\propto \nu^{-0.96 \pm 0.05}$, from which we infer $p\sim 3$, as typically found in radio SNe (e.g., \citealt{Chevalier06}). }
\label{Fig:VLAmodeling}
\end{figure}

\subsection{VLA Data Analysis}
We present in Fig. \ref{Fig:VLAmodeling} multi-band observations of \sn{} taken up to 278 days post explosion with the Karl G. Jansky Very Large Array (VLA, projects 16A-447 and 17A-167). The details of these data are given in Table \ref{Tab:radio}. We used standard phase referencing mode and the standard flux density calibrators 3C48 and 3C286 were used to set the absolute flux density scale. The data were calibrated using the VLA pipeline in \textsc{CASA} version 5.4.1, and imaged in \textsc{CASA} \citep{McMullin2007} following standard routines. We used Briggs weighting with a robust parameter of one to image. In the epochs where \sn{} was sufficiently bright, we performed phase-only self-calibration on the target. We subsequently fitted the sources in the image plane using the Python Blob Detector and Source Finder (\textsc{PyBDSF}, \citealt{Mohan15}). The uncertainties listed in Table \ref{Tab:radio} take into consideration the errors on the fit and a 5\% uncertainty on the absolute flux density scale. The flux density evolution of \sn{} at $\sim$8.5 GHz is presented in Fig. \ref{Fig:xraysALL} (left panel), together with a comparison with other SESNe and GRBs at the same frequency.
\subsection{Inferences on the Progenitor Properties and Mass-loss History from Radio Observations}
\label{SubSec:radiomodeling}
Radio emission in type Ibc SNe is well explained as synchrotron emission from relativistic electrons with a power-law distribution of 
Lorentz factors $\gamma$ ($N_e(\gamma)\propto \gamma^{-p}$) that gyrate in shock amplified magnetic fields (e.g., \citealt{Chevalier06}). 
\sn{} shows the characteristic ``bell-shaped'' spectrum of radio sources dominated by synchrotron self-absorption (SSA), with spectral peak flux $F_{pk}\propto t^{-0.31 \pm 0.02}$ and peak frequency $\nu_{pk}\propto t^{-0.97\pm0.02}$. By fitting a broken power-law to the radio data of \sn{} we find that the optically thin part of the spectrum $F_{\nu}\propto \nu^{-(p-1)/2}$ scales as $F_{\nu}\propto \nu^{-0.96 \pm 0.05}$, which implies $p\sim 3$, as typically found in radio SNe (e.g., \citealt{Chevalier06}). For the optically thick part of the spectrum our fits indicate $F_{\nu}\propto \nu^{2.4 \pm 0.1}$, consistent with the SSA expectation $F_{\nu}\propto \nu^{5/2}$. We find no evidence for free-free external absorption (e.g., \citealt{Weiler02}), which would cause the optically thick spectrum to be steeper than $F_{\nu}\propto \nu^{5/2}$. 

Using the SSA formalism by \cite{Chevalier98}, the best fitting $F_{pk}(t)$ and $\nu_{pk}(t)$ above translate into a constrain on the 
outer shock radius evolution $R_{sh}(t)$, magnetic field $B(R)$ and circumstellar density profile $\rho_{CMS}(R)$. We find evidence for a slightly decelerating blastwave with $R_{sh}(t)\propto t^{0.82\pm0.02}$ and $B(R)\propto R^{-1.14 \pm 0.03}$ propagating into a medium with density profile $\rho_{CSM}(R)\propto R^{-1.84 \pm 0.04}$. The inferred $B(R)$ profile is steeper than the $B(R)\propto R^{-1}$ scaling typically observed in H-stripped SNe (e.g., \citealt{Horesh13}) and causes the observed decay of $F_{pk}(t)$ with time. Normal non-decelerating type Ibc SNe typically show a constant $F_{pk}(t)$ \citep{Chevalier98}. The inferred $\rho_{CSM}(R)\propto R^{-1.84 \pm 0.04}$ is slightly flatter than a pure wind density profile $\rho_{wind}\propto R^{-2}$, which implies an \emph{increasing} effective mass-loss with radius $\dot M_{eff}\propto R^2 \rho_{CSM}\propto R^{0.16 \pm 0.04}$. We find $\dot M_{eff}(R_2)\sim2\times\dot M_{eff}(R_1)$, where $R_1 \sim4\times 10^{16}\,\rm{cm}$ is the blast wave radius at $10.5$ days and $R_2\sim10^{17}\,\rm{cm}$ is the blast wave radius at the end of the radio monitoring presented here, at $\delta t\sim 280$ days. For an assumed wind velocity $v_w=1000\,\rm{km\,s^{-1}}$ (appropriate for compact massive stars like WRs; \citealt{Crowther07}), these results imply that the stellar progenitor of \sn{} experienced a phase of enhanced mass-loss $\ge30$ yrs before collapse. We estimate that $\sim30$ yrs before death, the stellar progenitor of \sn{} was loosing twice the amount of material per unit time compared to $\sim$10 yrs before stellar demise.

According to the self-similar solutions by \cite{Chevalier82} the interaction of a steep SN outer ejecta profile $\rho_{SN}\propto R^{-n}$ with a shallower medium with $\rho_{CSM}\propto R^{-s}$ produces an interaction region that expands as $R_{sh}\propto t^{m}$ with $m=(n-3)/(n-s)$. For \sn, the inferred $R_{sh}(t)\propto t^{0.82\pm0.02}$ and $s= -1.84 \pm 0.04$ thus imply $n=8.2 \pm 0.7$. This result is consistent with the theoretical calculations of the post-explosion outer-ejecta density profiles of compact stars (e.g., WRs), for which \cite{Matzner99} find $n\sim10$. Extended red supergiants can have steeper outer density gradients \citep[e.g. $\gtrsim20$;][]{Fransson96}. We conclude that radio observations of \sn{} favor a compact progenitor star at the time of collapse.

All the considerations above do not depend on the assumed shock microphysical parameters $\epsilon_B$ and $\epsilon_e$ (i.e., the fraction of post-shock energy in magnetic fields and electrons, respectively). 
Below we provide the best-fitting values of the shock radius $R_{sh}$, internal energy $U$, magnetic field $B$ and effective mass loss $\dot M_{eff}$ at a given reference epoch under the assumption of equipartition of energy between electrons, protons and $B$ (i.e., $\epsilon_B=\epsilon_e=0.33$). 
Following \cite{Chevalier98}, we find:
\begin{equation}
\label{Eq:Bfield}
B(10.5\,\rm{d})=(4.0\pm0.2) \Big(\frac{\epsilon_e}{0.33}\Big)^{-\frac{4}{19}} 
\Big(\frac{\epsilon_B}{0.33}\Big)^{+4/19}\,\,\rm{G},
\end{equation}
\begin{equation}
\label{Eq:Radius}
R_{sh}(10.5\,\rm{d})=(3.1 \pm 0.1)\times 10^{15} \Big(\frac{\epsilon_e}{0.33}\Big)^{-1/19} \Big(\frac{\epsilon_B}{0.33}\Big)^{+1/19}\,\,\rm{cm}.
\end{equation}
The outer shock radius of Eq. \ref{Eq:Radius} does not strongly depend on the assumed microphysical parameter values. From Eq. \ref{Eq:Radius}, we can thus derive a solid estimate of the average SN shock velocity at $t=10.5$ days $v_{sh}\sim 0.15c$. This value is similar to normal type Ibc SNe (e.g., \citealt{Chevalier06}) and different from GRB-SNe and relativistic SNe, which show evidence for ultra-relativistic and mildly relativistic outflows \citep{Soderberg10,Margutti14b,Chakraborti15}. The effective mass-loss is:
\begin{equation}
\label{Eq:masslossRadio}
\dot M_{eff}(10.5\,\rm{d})=(3.6\pm0.3)\times 10^{-5} \Big(\frac{\epsilon_e}{0.33}\Big)^{-8/19} \Big(\frac{\epsilon_B}{0.33}\Big)^{-11/19} \,\,\rm{M_{\sun}yr^{-1}},
\end{equation}
and the shock internal energy is:
\begin{equation}
U(10.5\,\rm{d})=(1.1\pm 0.1)\times 10^{47}\Big(\frac{\epsilon_e}{0.33}\Big)^{-11/19} \Big(\frac{\epsilon_B}{0.33}\Big)^{-8/19}\,\,\rm{erg}.
\label{Eq:U}
\end{equation}
Under the assumption of equipartition, the internal energy value $U(10.5\,\rm{d})=(1.1\pm 0.1)\times 10^{47}\,\rm{erg}$ sets a lower limit on the true internal energy of the system at $t=10.5$ d, and on the kinetic energy of the radio emitting material. $U$ increases with time, as the shock decelerates and more kinetic energy of the shock wave is converted into internal energy. At $t\sim 280$ d we measure $v_{sh}\sim0.06\pm0.01$ c and $U(280\,\rm{d})=(7.6\pm 0.9)\times 10^{47}$ erg (in equipartition), which places \sn{} among energetic shocks from normal H-stripped SNe (Fig. 2 in \citealt{Margutti14b}).

Realistic values of $\epsilon_e$ and $\epsilon_B$ in SN shocks are likely $<0.33$, implying that both the equipartition $\dot M_{eff}$ and $U$ are lower limits on the true values of the system. 
For comparison, for more realistic values of $\epsilon_e=0.1$ and $\epsilon_B=0.01$ \citep[typical values for relativistic shocks;][]{Sironi2011}, we infer $\dot M_{eff}(10.5\,\rm{d})= (4.5\pm0.4)\times 10^{-4}\,\rm{M_{\odot}yr^{-1}}$ and $U(280\,\rm{d})=(6.7\pm 0.8)\times 10^{48}$ erg. 
Recent kinetic simulations of trans-relativistic shocks suggest values of $\epsilon_B\sim 0.01$ and $\epsilon_e\gtrsim 10^{-3}$ \citep{Park2015,Crumley2019}.
An $\epsilon_e\sim 3\times 10^{-3}$ would imply a very energetic explosion, with $U(280\,\rm{d})=(5.1\pm 0.6)\times 10^{49}$ erg, however it also yields an unrealistic $\dot M_{eff}(10.5\,\rm{d})= (2.0\pm0.2)\times 10^{-3}\,\rm{M_{\odot}yr^{-1}}$. 
Similar values of mass-loss are more typical of progenitor stars of Type IIn SNe and are likely too high for \sn{}.
In general, the theory of particle acceleration at strong trans-relativistic shocks not only does not explain large values of $\epsilon_e$, but also does not produce the typical $p\sim 3$ often inferred in radio SNe.
The explanation of \citealt{Chevalier06} invokes shocks modified by the dynamical backreaction of the accelerated particles, which was thought to lead to concave spectra, steeper than $E^{-2}$ below a few GeV, but such an argument is at odds with observations of Galactic SN remnants \citep{Caprioli12}.

A robust upper limit on the effective mass-loss can be inferred from the lack of free-free external absorption in the radio spectra. Indeed, the absence of a low-frequency cut-off can be used to constrain the environment density independently from the shock microphysics. From \cite{Weiler02}, the free-free optical depth of unshocked ionized gas in a wind density profile is: 
\begin{equation}
\tau_{\rm ff} \simeq \frac{\alpha_{\rm ff}r}{3} \approx \newline 10\left(\frac{\nu}{10{\rm GHz}}\right)^{-2}\left(\frac{T_{\rm g}}{10^{4}{\rm K}}\right)^{-3/2} \dot M_{-3}^2 \left(\frac{v_{sh}}{0.1c}\right)^{-3}t_{\rm wk}^{-3}\,\,,
\end{equation}
where $\dot M$ is in units of $10^{-3}\,\rm{M_{\sun}yr^{-1}}$ for $v_{w}=1000\,\rm{km\,s^{-1}}$, $T_{\rm g}$ is the temperature of the gas, normalized to a value $T_{\rm g} \gtrsim 10^{4}$ K typical of photoionized gas, and time is units of 1 week. Furthermore, we used $\kappa_{\rm es} = 0.38$ cm$^{2}$ g$^{-1}$ for fully ionized solar-composition ejecta and $\alpha_{\rm ff} \approx 0.03 n_w^{2}\nu^{-2}T_{\rm g}^{-3/2}$ cm$^{-1}$ as the free-free absorption coefficient. The lack of evidence for free-free absorption at 10.5 days at $\nu=5.9$ GHz, and at 45.5 days at $\nu=3$ GHz demands $\tau_{ff}\ll1$, which translates into $\dot M<10^{-3}\,\rm{M_{\sun}yr^{-1}}$ for $v_{w}=1000\,\rm{km\,s^{-1}}$.

\section{X-rays} \label{sec:XRT}

\subsection{Swift-XRT and XMM-Newton Data Analysis}
\label{SubSec:Xraydata}

\begin{figure*}
\vskip -0.0 true cm
\centering
\includegraphics[width=0.48\textwidth]{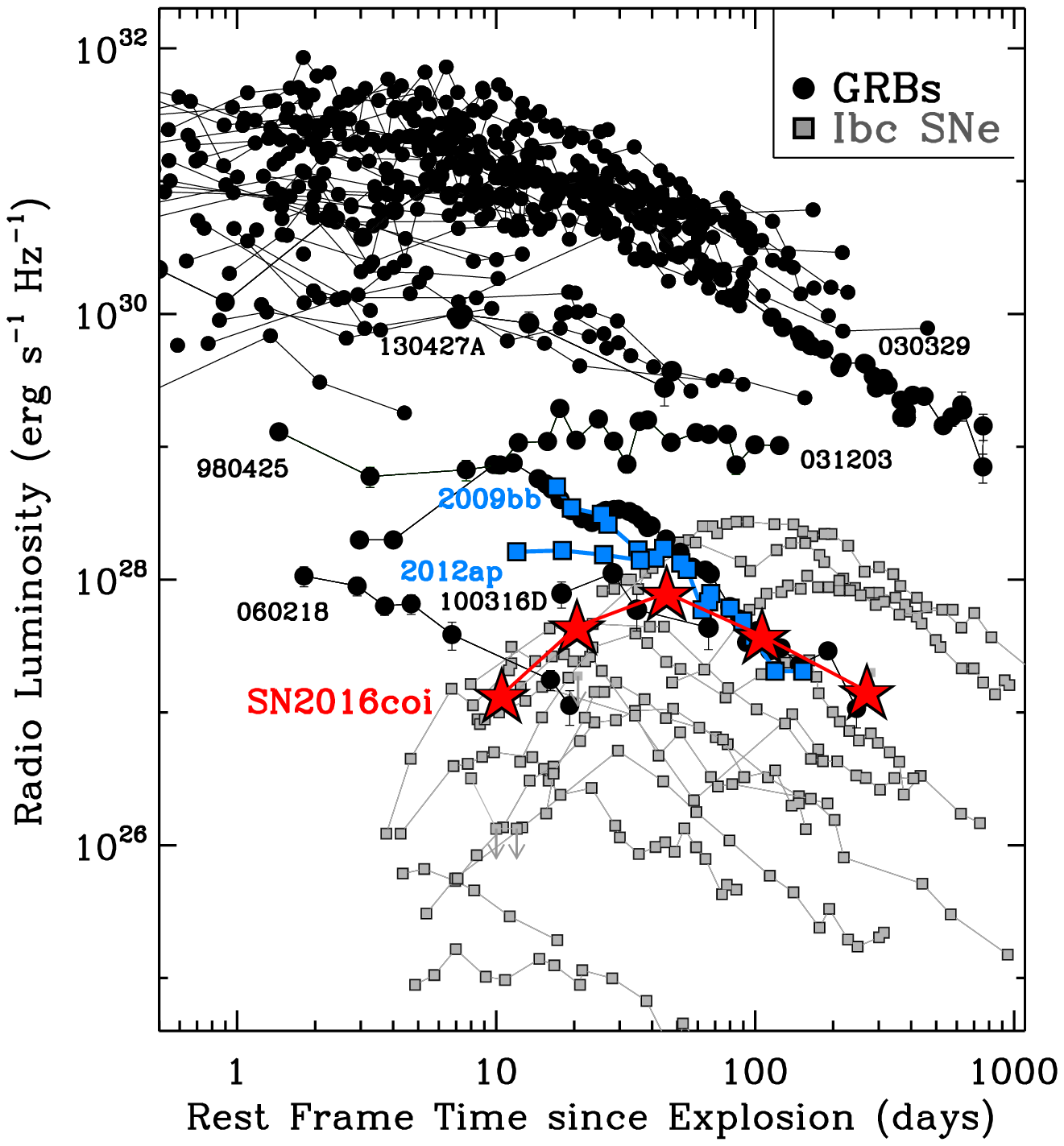}
\includegraphics[width=0.48\textwidth]{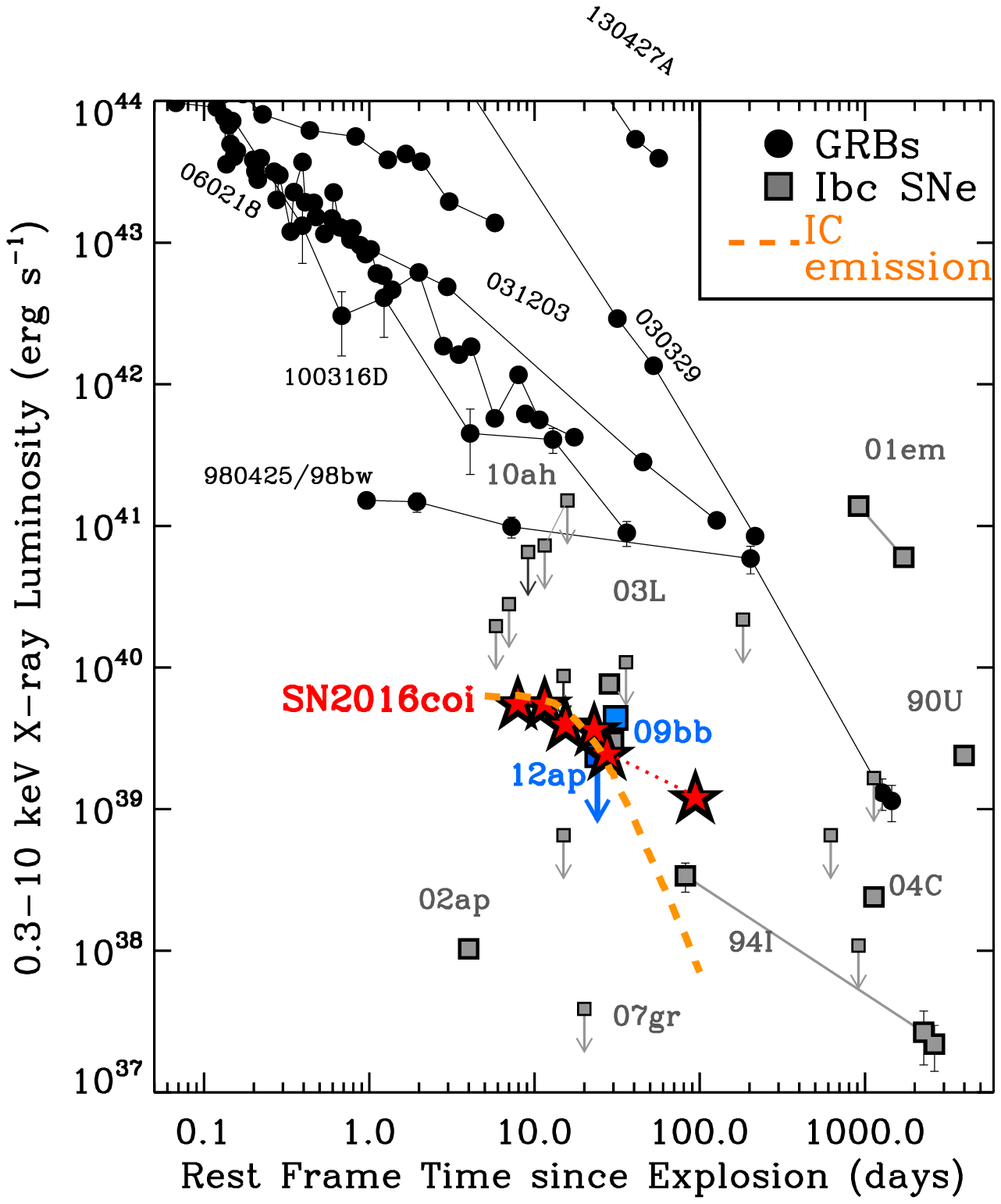}
\caption{Radio $\sim$8.5 GHz (left panel) and X-ray (right panel) emission from \sn{} (red stars) in the context of normal H-stripped SNe (grey squares), relativistic SNe (blue squares) and GRBs (black filled circles). While being significantly fainter than GRB-SNe, \sn{} competes in X-ray luminosity with relativistic SNe and it is significantly more luminous than the BL-Ic SN\,2002ap. The radiation from \sn{} is well explained by synchrotron emission at radio wavelengths at all times, while the X-rays are dominated by Inverse Compton (IC) emission (dashed orange line).
At late times the IC scattering model underestimates the observed X-ray flux of \sn{}, suggesting additional contributions from other mechanisms. 
 References: \cite{Immler02,Pooley04,Soria04,Soderberg05,Perna08,Soderberg10,Chakraborti11b,Chandra12,Horesh13,Margutti13,Chakraborti14,Corsi14,Margutti14b}.}
\label{Fig:xraysALL}
\end{figure*}

The X-Ray Telescope \citep[XRT;][]{Burrows05}, on board the Neil Gehrels \textit{Swift} Observatory, started observing \sn{} on May 27, 2016 ($\delta t\sim2$ days post explosion) until April 17, 2017 ($\delta t\sim326$ days), for a total exposure time of 94.4 ks. \emph{Swift}-XRT data have been analyzed using the latest HEAsoft release v6.22 and corresponding calibration files. Standard filtering and screening criteria have been applied (see \citealt{Margutti13} for details). An X-ray source is clearly detected at the location of \sn{} until $\delta t\sim100$ days post explosion. The X-ray source is located $\sim30''$ from the host galaxy nucleus (which is not detected by \emph{Swift}-XRT and does not represent a source of contaminating X-ray emission; see Fig. \ref{Fig:rgb}, bottom panel) and shows a fading behavior with time, from which we conclude that the detected X-ray emission is physically associated with \sn{}. The spectrum can be fit with an absorbed power-law spectral model with best fitting photon index $\Gamma= 1.78 \pm 0.18$. We find no evidence for intrinsic absorption and we place a $3\sigma$ limit for the neutral hydrogen absorption column $NH_{i}<0.4\times 10^{22}\,\rm{cm^{-2}}$. The Galactic $NH_{i}$ in the direction of \sn{} is $NH_{mw}=0.056\times 10^{22}\,\rm{cm^{-2}}$ \citep{Kalberla05}. For this spectrum, the 0.3-10 keV count-to-flux conversion factor is $\sim 4.22\times 10^{-11}\,\rm{erg\,s^{-1}cm^{-2}ct^{-1}}$ (unabsorbed). The \emph{Swift}-XRT count-rate and flux-calibrated light-curve is reported in Table \ref{Tab:xrays} and shown in Fig. \ref{Fig:xraysALL} (right panel).

We started deep X-ray observations of the field of \sn{} with XMM-Newton on June 6, 2016 (PI Margutti). We obtained two epochs of observations at $\delta t\sim 11.5$ days (exposure time of 29 ks, observation ID 0782420201) and $\delta t\sim 27.5$ days (exposure of 28 ks, observation ID 0782420301). XMM data have been analyzed with SAS (v15.0). 
The first observation was heavily affected by proton flaring and the net exposure time of the EPIC-pn camera after filtering out the intervals of high background was reduced to 1.1 ks, whereas for the second epoch we have 13.4 ks net exposure time. \sn{} is clearly detected by all three cameras in both epochs. The inferred EPIC-pn count-rate is $(4.0\pm0.7)\times 10^{-2}\,\rm{ct\,s^{-1}}$ and $(1.7\pm0.2)\times 10^{-2}\,\rm{ct\,s^{-1}}$ (0.3-10 keV) for the first and second epoch, respectively. A spectrum extracted from the first (second) epoch can be fitted with an absorbed power-law model with $\Gamma=1.9\pm 0.3$ ($\Gamma=1.8\pm 0.3$). The corresponding flux is $\sim 1.1\times 10^{-13}\,\rm{erg\,s^{-1}cm^{-2}}$ and $\sim 0.6\times 10^{-13}\,\rm{erg\,s^{-1}cm^{-2}}$ for the first and the second epoch, respectively, consistent with the results from our \emph{Swift}-XRT monitoring (Table \ref{Tab:xrays} and Fig. \ref{Fig:xraysALL}). 

We do not find evidence for significant spectral evolution. From a joint fit of \emph{Swift}-XRT and XMM data we find a best-fitting $\Gamma=1.80\pm0.10$ and $NH_{i}<0.17\times 10^{22}\,\rm{cm^{-2}}$.

\subsection{Inferences on the Mass-loss History of the Stellar Progenitor from X-ray Observations} 
\label{Sec:massloss}
In normal H-poor SNe, the early time ($\delta t\lesssim 30$ days) X-ray emission is expected to be dominated by Inverse Compton (IC) scattering of optical photospheric photons by relativistic electrons accelerated at the shock fronts (e.g., \citealt{Bjornsson04,Chevalier06}). The non-thermal X-ray spectrum of \sn{} with $\Gamma\sim 2$ and lacking evidence for intrinsic absorption is consistent with this expectation. Adopting the formalism by \cite{Margutti12}, the IC emission depends on: (i) density profile of the SN ejecta $\rho_{ej}$; (ii) properties of the electron distribution responsible for the up-scattering $N_e(\gamma)$; (iii) blastwave velocity, which, in turns, depends on the circumstellar medium (CSM) density and explosion's parameters (kinetic energy $E_{k}$ and ejecta mass $M_{ej}$); (iv) optical bolometric luminosity of the SN (from \S \ref{sec:Lbol}), which is the ultimate source of photons that are upscattered to X-ray energies $L_{X,IC}\propto L_{bol}$. We parametrize the CSM density as a wind medium $\rho_{CSM}=\dot M/4\pi v_w R^{2}$ (where $v_w$ is the progenitor wind and $\dot M$ is the mass-loss rate) and we use $\rho_{ej}\propto R^{-n}$ with $n\sim10$, as appropriate for SNe with compact progenitors (e.g., \citealt{Matzner99,Chevalier00}), consistently with the results from the modeling of radio data in \S \ref{SubSec:radiomodeling}. We further assume a power-law distribution of electrons $N_e(\gamma)\propto \gamma ^{-p}$ with $p\sim3$ and $\epsilon_e=0.1$ for consistency with the modeling of other SNe (e.g., \citealt{Chevalier06}). 

\begin{figure}
\vskip -0.0 true cm
\centering
\includegraphics[width=\columnwidth]{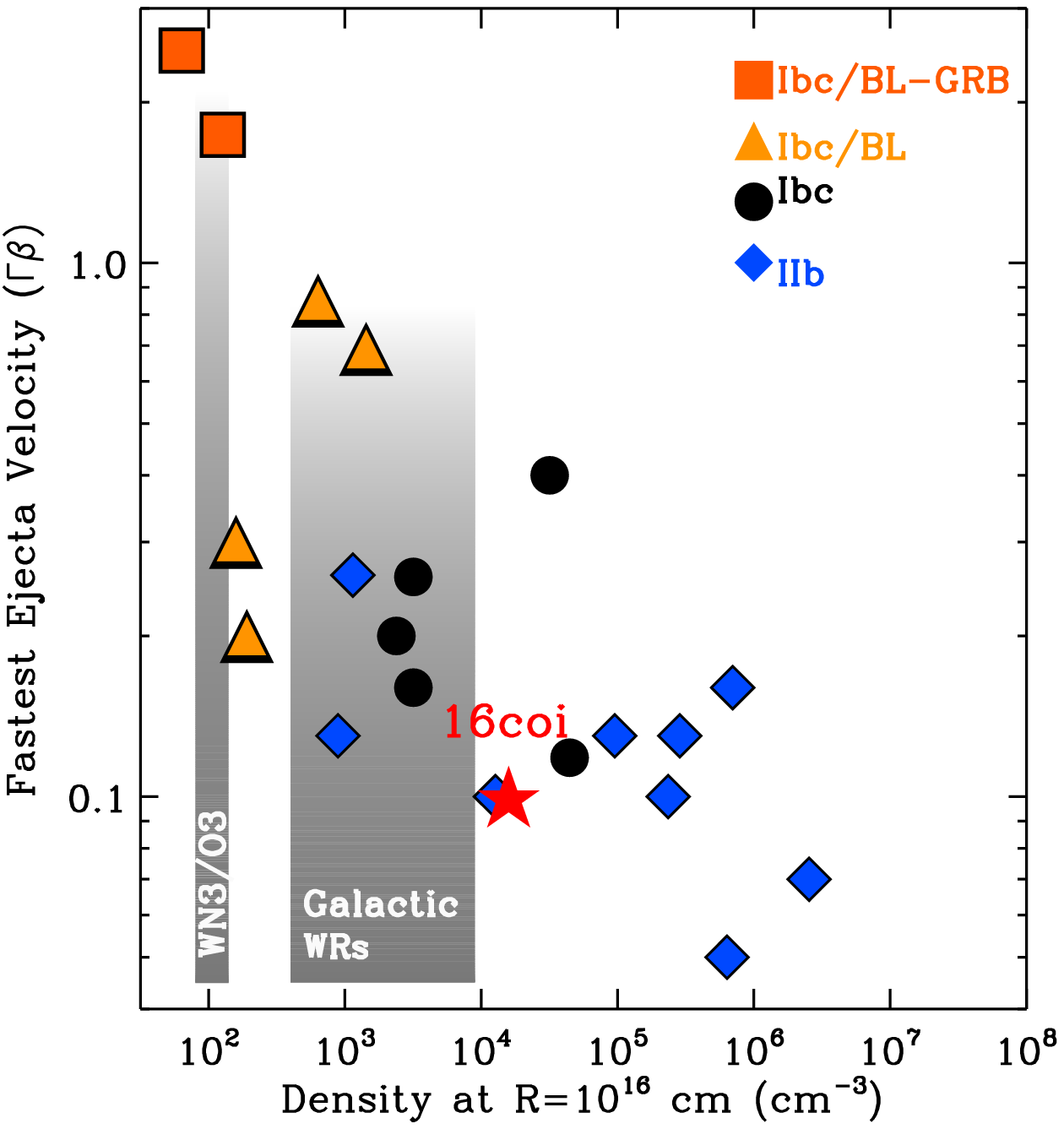}
\caption{Fastest ejecta velocity in the explosion vs. environmental number density of \sn{} in the context of H-stripped core-collapse SNe. Type IIb SNe (blue diamonds) explode in the densest environments, while SN that accompany GRBs are associated with the lowest density environments (orange squares). Normal type Ibc SN are shown with black filled circles. SN with broad features in their spectra (Ibc/BL in the plot, orange triangles) also tend to be associated with low density media. An exception to this behavior is \sn{}, which exploded in a dense environment (red star). For \sn{} we show here the equipartition number density. The true number density in the environment of \sn{} is $\sim10$ times the equipartition value if $\epsilon_B= 0.01$. Grey shaded regions: density in the environments of WRs and the recently discovered new type of WR stars WN3/O3 (\citealt{deJager88,Marshall04,vanLoon05,Crowther07,Massey15}). References: \cite{vanDyk94,Fransson98,Berger02,Weiler02,Ryder04,Soderberg05,Chevalier06,Soderberg06b,Soderberg06c,Soderberg08,Roming09,Soderberg10,Soderberg10b,Krauss12,Milisavljevic13,Margutti14b,Kamble14,Corsi14,Chakraborti15,Drout16,Kamble16,Margutti17}. }
\label{Fig:massloss}
\end{figure}

Considering the range of explosion parameters of \S \ref{sec:Lbol} and assuming a wind velocity of $v_w=1000\,\rm{km\,s^{-1}}$, for a shock velocity of $v_{sh}\sim0.1$ c (\S \ref{SubSec:radiomodeling}) we infer a mass-loss rate of $\dot M\sim (1-2)\times 10^{-4}\,\rm{M_{\sun}yr^{-1}}$.
Our X-ray analysis thus provides independent evidence that \sn{} exploded in a dense environment when compared to type Ic-BL SNe (Fig. \ref{Fig:massloss}). This suggests that the stellar progenitor of \sn{} experienced significant mass loss in the last years before core-collapse. This result is independent of $\epsilon_B$. A comparison to Eq. \ref{Eq:masslossRadio} suggests that for $\epsilon_e=0.1$ $\epsilon_B\le0.1$.

The right panel of Fig. \ref{Fig:xraysALL} clearly shows that IC (dashed orange line) fails to reproduce the bright X-ray emission at $\sim$ 100 days by a large factor. At these times, synchrotron emission is expected to dominate (e.g., \citealt{Chevalier06}). The extrapolation of the optically thin $F_{\nu}\propto \nu^{-0.96}$ radio spectrum to the X-ray band under-predicts the observed X-ray emission by a large factor $\sim30$. The discrepancy between the extrapolation of synchrotron spectrum that best fits the radio and the observed X-ray data is even larger when we consider that the 
synchrotron cooling frequency $\nu_c=18\pi m_{e} c q/(t^2\,B^3\,\sigma_T^2)$ is $\nu_c\sim 10^{11}-10^{12}\,\rm{Hz}$ at $\sim 100$ days using $B$ from Eq. \ref{Eq:Bfield} and $\epsilon_B=0.01-0.1$ (where $q$ is the electron charge and $m_e$ is the electron mass). Above $\nu_c$ the flux density steepens as $F_{\nu}\propto \nu^{-p/2}$, leading to an even lower expected X-ray flux.
The conclusion is that the late time $t\ge100$ days X-ray emission from \sn{} is too luminous to be explained within the standard framework of synchrotron radiation from a population of electrons accelerated into a simple power-law distribution $N_e(\gamma)\propto \gamma^{-p}$. 

The problem of having very luminous X-ray emission from H-stripped core-collapse SNe at late times is not new and was explored in detail by \cite{Chevalier06}. These authors favor an interpretation where the particle spectrum is modified and becomes flatter for $\gamma\ge1000$. The net effect is an increase of the X-ray synchrotron emission, while the effect on the radio synchrotron emission is minor (see their Fig. 1). At the time of writing it is unclear what physical effect might produce this shape of the particle spectrum, as the cosmic-ray dominated shocks invoked by \cite{Chevalier06} have not been confirmed by recent particle-in-cell (PIC) simulations \citep{Park2015}. 
We end by noting that at the large mass-loss rates inferred for \sn{} in the case of deviation from equipartition $\dot M_{eff}\sim5\times 10^{-4}\rm{M_{\sun}\,yr^{-1}}$, the X-rays are likely to receive a contribution from free-free emission. From \cite{Chevalier06}, their Eq. 30, for this mass-loss rate we estimate $L_{x,ff}\sim5\times 10^{38}\,\rm{erg\,s^{-1}}$ at $t\sim100$ days, which is a factor $\sim2$ lower than the observed X-ray emission at this epoch.

\section{Search for shock breakout emission at high energies} \label{sec:SBO}
\label{Sec:SBO}
For compact massive H-stripped stars that are progenitors of (some) type Ibc SNe, the very first electromagnetic signal able to escape from the explosion site and reach the observer (i.e., the breakout pulse) is expected to peak at X-ray and $\gamma$-ray energies (e.g., SN\,2008D; \citealt{Soderberg08}). We searched for a high-energy pulse associated with the shock breakout of \sn{} using data collected by the InterPlanetary Network (IPN), which includes Mars Odyssey, Konus-\emph{Wind}, RHESSI, INTEGRAL (SPI-ACS; SPectrometer on INTEGRAL-Anti-Coincidence System), \emph{Swift}-BAT (Burst Alert Telescope) and \emph{Fermi}-GBM (Gamma-Ray Burst Monitor). The IPN observes the entire sky with temporal duty cycle $\sim100$\% when all the experiments are considered. A total of 6 bursts were detected by the spacecraft of the IPN between May 22.4, 2016 and May 25.6, 2016, which covers the most likely explosion date window May $23.9 \pm 1.5$ days that we inferred in \S \ref{sec:Lbol}. None has a localization region consistent with the position of \sn{}.
We thus conclude that there is no evidence for a SN-associated shock breakout pulse down to the IPN sensitivity threshold with fluence $F_\gamma \sim 6\times 10^{-7}\,\rm{erg\,cm^{-2}}$ ($E_{\gamma}\sim 2\times 10^{46}\,\rm{erg}$ at the distance of \sn{}).

\subsection{Comparison to Breakout Models}
Following \cite{Katz12} (and references therein), the expected shock breakout energy is $E_{BO}=8 \pi R_*^{2} v_0 c \kappa^{-1}$ (their Eq. 40), where $R_*$ is the stellar radius, $\kappa$ is the opacity and $v_0$ is the shock velocity at breakout. For \sn{} we assume $\kappa\sim0.4\,\rm{cm^2\, g^{-1}}$ and adopt $v_0 \approx 0.3\,\rm{c}$ (i.e., a shock velocity at breakout similar to the maximum ejecta velocity as inferred from the X-ray observations, see \citealt{Katz12}, their Eq. 25). For compact progenitors like WR stars with $R_*=10^{11}\,\rm{cm}$ we find $E_{BO}\approx 10^{44}\,\rm{erg}$, significantly below the IPN sensitivity. \emph{Fermi}-GBM and \emph{Swift}-BAT are more sensitive and reach fluence limits of $\sim4\times 10^{-8}\,\rm{erg\,cm^{-2}}$ and $\sim6\times 10^{-9}\,\rm{erg\,cm^{-2}}$, respectively, corresponding to $E_{\gamma}\sim 2\times 10^{45}\,\rm{erg}$ and $E_{\gamma}\sim 2\times 10^{44}\,\rm{erg}$. The \emph{Swift}-BAT threshold for detection is comparable to the expected $E_{BO}$. \emph{Swift}-BAT observes $\sim1/6$ of the sky with $\sim90$\% temporal duty cycle. It is thus possible that \emph{Swift}-BAT missed the breakout pulse, or that the breakout pulse lies just below the \emph{Swift}-BAT threshold of detection. For comparison the breakout pulse in SN\,2008D showed $L_x\sim10^{44}\,\rm{erg\,s^{-1}}$ (0.3-10 keV) at peak with a duration of $\sim5$ minutes.

More extended progenitors with radii $R_*=10^{13}\,\rm{cm}$ would lead to inferred $E_{BO}\approx 10^{48}\,\rm{erg}$. In this case however the spectrum of breakout pulse is expected to peak at lower frequencies $<1$ keV which are not probed by the hard X-ray/$\gamma$-ray observations presented here. A similar reasoning and conclusion apply if the radiation breakout occurred in a thick medium outside the star.

\section{Discussion} 
\label{sec:disc}
Our data analysis and modeling characterize \sn{} as an energetic H-stripped SN with (i) He in the ejecta, (ii) a broad bolometric light-curve, and (iii) luminous X-ray and radio emission. These three observables distinguish \sn{} from the rest of the population of known H-stripped SNe and directly map into properties of its progenitor star: a massive, well-mixed star that experienced substantial mass loss in the years preceding core-collapse. We discuss below the implications of these findings in the broader context of stellar progenitors of H-stripped SNe.

\subsection{Broad Bolometric Light-Curve and Nebular Spectroscopy Indicate a Massive Progenitor}
\label{subsec:broad LC}
Among H-stripped SNe, \sn{} shows one of the broadest bolometric light-curves (Fig. \ref{Fig:LCcomparison}), from which we infer $M_{ej}\sim5-7\,\rm{M_{\sun}}$ \citep[\S \ref{sec:Lbol};][]{Kumar18,Prentice17}. This value is larger than the typical ejecta mass $M_{ej}\sim2-3\,\rm{M_{\sun}}$ inferred for H-stripped SNe \citep[e.g.,][]{Drout2011,Bianco2014,Lyman2016,Taddia2018}.

Other type Ic SNe with broad light-curves are SNe 2004aw \citep{Taubenberger2006} and 2011bm \citep[the broadest light curve in Fig. \ref{Fig:LCcomparison};][]{Valenti2012}, for which the inferred ejecta mass is $3.5-8.0$ and $7-17$~\Msun{}, respectively.
Additionally, SN\,2004aw also displayed relatively high ejecta velocities ($v\sim12000$~km/s$^{-1}$) around maximum light, similar to \sn{}. Interestingly, a tentative identification of He in the ejecta of SN\,2004aw has also been reported based on NIR spectroscopy \citep{Taubenberger2006}.

Nebular spectroscopy of \sn{} provides additional constraints on the mass of its stellar progenitor. The relative abundances of different elements in a stellar envelope depend on the core mass. In particular, in the models by \cite{Fransson1989} the ratio of the integrated fluxes of the [\ion{Ca}{ii}] \lam\lam7291,7324 doublet and the [\ion{O}{i}] \lam\lam6300,6364 doublet can be used as an indicator of the progenitor core mass, 
with lower values signifying of more massive cores. This fact mainly results from two factors: first, in the models by \cite{Fransson1989} the relative abundance Ca/O is lower for progenitors with a more massive core, producing a lower [\ion{Ca}{ii}]/[\ion{O}{i}] flux ratio; second, in stars with a smaller core mass and with a stratified envelope like the ones in \cite{Fransson1989} models, the Ca is more mixed within the oxygen layers. Since Ca is a significantly more efficient coolant, this translates into more prominent [\ion{Ca}{ii}]/[\ion{O}{i}] ratio. 
However, several other factors can play a role in determining the observed [\ion{Ca}{ii}]/[\ion{O}{i}] flux ratio. \cite{Fransson1989}, for example, showed that higher densities of the ejecta (i.e., lower kinetic energies) would also lead to lower [\ion{Ca}{ii}]/[\ion{O}{i}] flux ratios.
Additionally, a high degree of mixing of the entire stellar envelope, where oxygen is more centrally located, also leads to a lower [\ion{Ca}{ii}]/[\ion{O}{i}] flux ratio (as in this case cooling through oxygen would act as a competitor to calcium in the inner envelope). These factors make the observed [\ion{Ca}{ii}]/[\ion{O}{i}] flux ratio a non-monotonic tracer of the stellar core-mass.

\begin{figure}
\vskip -0.0 true cm
\centering
\includegraphics[width=\columnwidth]{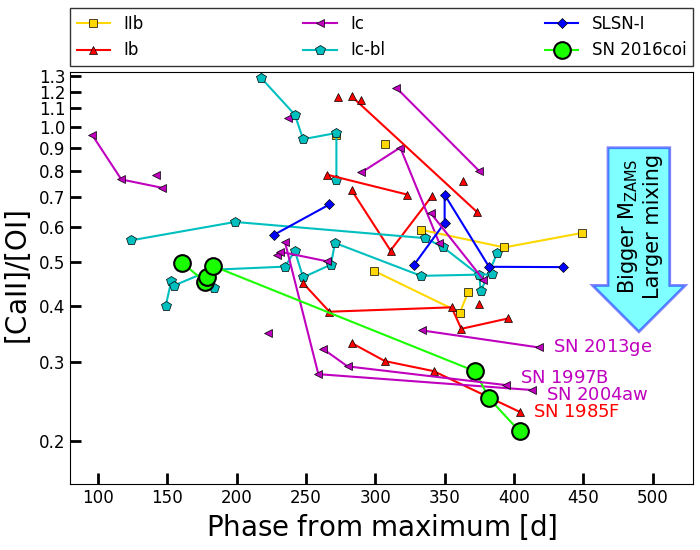}
\caption{Comparison between the evolution of the [\ion{Ca}{ii}] \lam\lam7291,7324 to [\ion{O}{i}] \lam\lam6300,6364 ratio for \sn{} and other type Ic and BL-Ic SNe. \sn{} is characterised by low [\ion{Ca}{ii}]/[\ion{O}{i}] ratio, suggesting a high \Mzams{} progenitor, with a high level of mixing at the time of explosion. Other events with a low ratio are SNe 1985F, 1997B and 2004aw. Measurements were performed depending on availability of late-time spectra retrievable from the literature \citep{Gaskell1986,Filippenko1986,Filippenko1995,Barbon1999,Patat2001,Foley2003,Elmhamdi2004,Taubenberger2006,Tanaka09,Taubenberger2009,Milisavljevic2010,Valenti2011,Silverman2012,Valenti2012,Benetti2011,Ben-Ami2014,Modjaz2014,Ergon2015,Kuncarayakti2015,Milisavljevic15c,Milisavljevic15,Smartt2015,Drout16,Nicholl2016,Kangas2017,Taddia2019}.}
\label{Fig:Ca_O}
\end{figure}

With these caveats in mind, the [\ion{Ca}{ii}]/[\ion{O}{i}] flux ratio has been used in the past as a diagnostic for the progenitor star \Mzams{} in core-collapse SNe \citep[e.g.,][]{Fransson1987,Fransson1989,Elmhamdi2011,Kuncarayakti2015}. We therefore compute this ratio as a function of time in \sn{} and show the results in Fig. \ref{Fig:Ca_O}. To build a homogeneous comparison sample, we retrieved late-time spectra of stripped-envelope SNe available from the literature\footnote{The spectra were retrieved from WISeREP \citep[\url{https://wiserep.weizmann.ac.il}][]{Yaron2012} and the OSC \citep[\url{https://sne.space};][]{Guillochon2017}.}. We select all the SNe with observations at $t>$100 days after explosion, and with at least one spectrum covering both the [\ion{O}{i}] and the [\ion{Ca}{ii}] region. We focus on those SNe with a [\ion{Ca}{ii}]/[\ion{O}{i}] ratio below 1.3. Our sample comprises 83 spectra from 29 different SNe. We then fit the line doublets with two Gaussian profiles (see Fig. \ref{fig:neb}). The position of the first centroid is kept as a free parameter, but the separation between the two lines of each doublet is kept fixed at the expected value.

From Fig. \ref{Fig:Ca_O} it is clear that at very late phases \sn{} occupies the lower part of the plot, and at $t>400$~d has the lowest [\ion{Ca}{ii}]/[\ion{O}{i}] ratio, close to $\sim0.2$. For reference, \cite{Fransson1989} found ratios of $\sim0.6$ and $\sim5.6$ for their 8~\Msun{} and 4~\Msun{} He-core progenitor models, respectively. This analysis independently supports the idea that \sn{} originated from a stellar progenitor with larger mass than the average progenitor of H-stripped core-collapse SNe. However, as described above, other factors can contribute to the observed [\ion{Ca}{ii}]/[\ion{O}{i}] ratio, like mixing and ejecta densities. Indeed, the values measured by \cite{Fransson1989} would become $\sim0.3$ and $\sim1.6$ for the same progenitor models as above, with lower explosion kinetic energies ($\sim$8 times higher densities).

Other SNe with low flux ratios ($<0.3$) are the type Ib SN 1985F \citep{Schlegel1989,Elmhamdi2004}, and the two type Ic SNe 1997B (spectra retrieved from the Asiago Supernova Archive) and 2004aw \citep{Taubenberger2006}. 
Interestingly, at least two of these SNe also have broad light curves. The bolometric light-curve of SN\,2004aw is very similar to \sn{} (Fig. \ref{Fig:LCcomparison}), while SN\,1985F has an even broader light curve, with a $\Delta_{15}$ in $B$-band of 0.52~mag \citep{Tsvetkov1986} ($\Delta_{15}$=1.01~mag for \sn{} \citealt{Kumar18}). Unfortunately, SN\,1997B was discovered after peak. 
Consistent with the caveats above, some SNe with broad light curve have large [\ion{Ca}{ii}]/[\ion{O}{i}] ratio (e.g., SN\,2011bm, \citealt{Lyman2016}). 

Based on the estimated $M_{ej}\sim 4-7$~\Msun{} (\S \ref{sec:Lbol}), and assuming a fiducial mass for the remnant compact object between 1.5~\Msun{} (for a neutron star) and 3~\Msun{} (for a black hole), we estimate a mass of the C+O stellar progenitor of \sn{} at the time of collapse of $\sim$6-10~\Msun{}. Similar values have been inferred for the progenitor of SN\,2004aw, for which \cite{Mazzali2017} estimated a ZAMS mass of $\sim23-30$~\Msun.\footnote{The field of \sn{} was not serendipitously observed by HST before explosion, which prevents a constraining search for a progenitor star in pre-explosion images. The \Ha{} Galaxy Survey \citep{James2004} observed the field of \sn{} on June 6th, 2000. A compact source of \Ha{} emission is clearly detected $\sim4.9$\arcsec\, from the SN location, (yellow mark in Fig. \ref{Fig:rgb}). At the distance of \sn{}, this angular separation corresponds to a projected distance of 0.43 kpc. \cite{Prentice17} estimated 0.375 kpc, using however a shorter distance to the host galaxy (see Table \ref{tab:other_works}).} The actual value of the inferred ZAMS mass strongly dependends on the adopted mass loss prescriptions \citep[e.g.,][]{Smith2014}.

\subsection{Luminous Radio and X-ray Emission from Large Progenitor Mass-loss before Explosion}
\label{SubSec:MassLossDisc}

\begin{figure}
\vskip -0.0 true cm
\centering
\includegraphics[width=\columnwidth]{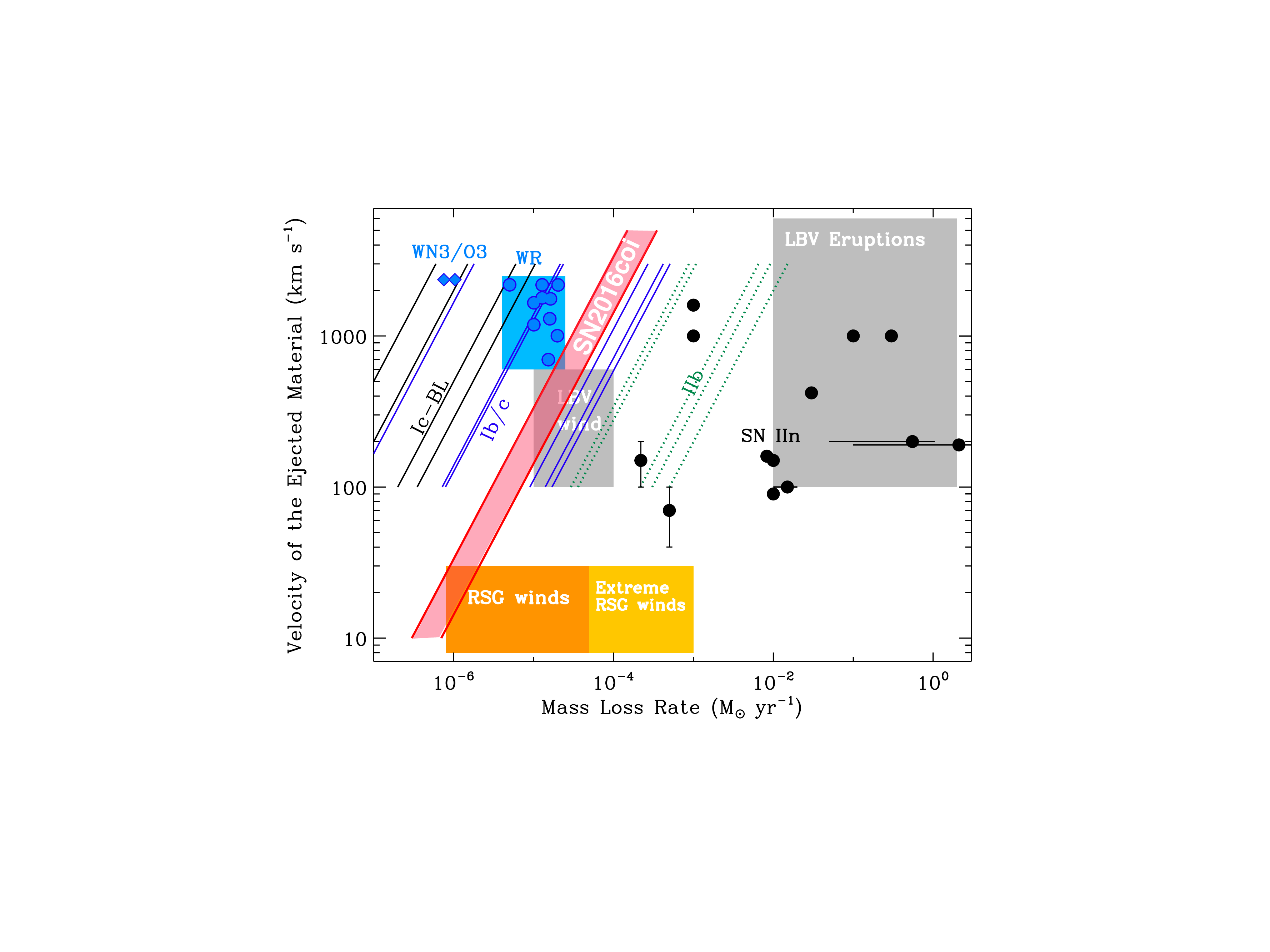}
\caption{Constraints on the recent mass-loss history of \sn{} (red shaded area) in the context of observed mass-loss rates and wind velocities in massive stars. Here we conservatively plot the equipartition $\dot M$. The true \sn{} $\dot M$ is $\sim10$ times larger if $\epsilon_B= 0.01$. Galactic WR stars from \citealt{Crowther07}, WN3/O3 stars from \citealt{Massey15}, red supergiants (RSGs) winds from \citealt{deJager88,Marshall04,vanLoon05}. Typical locations of Luminous Blue Variable (LBV) winds and eruptions are from \citealt{Smith2014} and \citealt{Smith06}. Black, blue and dotted green lines mark the sample of type Ic-BL, Ibc and IIb SNe from \cite{Drout16}. Inferred mass-loss rates for type-IIn SNe are from \cite{Kiewe12}.}
\label{Fig:massloss2}
\end{figure}

The recent mass-loss history of the progenitor star in the centuries leading up to the explosion can be constrained with radio and X-ray observations, which sample the emission originating from the interaction between the fastest SN ejecta and the CSM. The resulting luminosity mainly depends on the shock velocity and on the environment density, with faster shocks and denser environments powering the most luminous radio and X-ray displays.

We compare the properties of \sn{} that we inferred in \S \ref{SubSec:radiomodeling} and \ref{Sec:massloss} to a sample of H-stripped SNe in Fig. \ref{Fig:massloss}. SNe with fast ejecta velocities like Ic-BL SNe (orange squares and triangles in Fig. \ref{Fig:massloss}) tend to be associated with low-density environments, while type IIb SNe are located within the densest circumstellar media. From the radio we inferred a shock velocity of $v_{sh}\sim0.15c$ for \sn{}. This sub-relativistic shock can be caused either by a lower shock velocity at breakout, or by a denser-than-average environment surrounding the progenitor. In the latter case, the CSM was sculpted by a prolonged enhanced mass-loss phase of the stellar progenitor in the years before stellar death. The environment of \sn{} is among the densest in the sample of type Ibc SNe. Assuming equipartion of energy, from the radio data we obtained a lower limit for the mass-loss of \sn{} of $\dot M \sim 3-4 \times 10^{-5}$~\Msun{}yr$^{-1}$ (for $v_w=1000$~km s$^{-1}$). X-ray analysis pushed this value even higher, with $\dot M \sim 1-2 \times 10^{-4}$~\Msun{}yr$^{-1}$. Such a large mass-loss rate is consistent with those associated with extreme line-driven winds in WR stars \citep{Crowther07}, as we show in Fig. \ref{Fig:massloss}-\ref{Fig:massloss2}. This result is also supported by the radio modelling, which showed that the post-explosion density profile of the outer ejecta is consistent with having originated from a compact object like a WR star. The inferred $\dot M$, \emph{if} sustained for the entire $\sim10^{5}\rm{yr}$ duration of the WR phase, implies a total mass-loss of several $M_{\sun}$ possibly sufficient to strip the progenitor star of a large fraction of its helium envelope even in the absence of interactions with a binary companion \footnote{The measured mass-loss refers to the WR-phase of the progenitor, therefore little could be said about the process responsible for the stripping of its hydrogen envelope.}. This scenario is consistent with the indication of a massive stellar progenitor of \S \ref{subsec:broad LC}.

\subsection{He Spectral Features as a Signature of Asymmetries?}
\label{SubSec:Hedisc}
The presence of He in \sn{}
is supported by the comparison with the velocity profile of \ion{Si}{ii} \lam6355 (\S \ref{sec: class}), and by the detailed spectral modeling of \cite{Prentice17}. \ion{He}{i} lines are formed through non-thermal excitation and ionization \citep{Lucy1991}, for example by $\gamma$-rays produced by the radioactive decay of \Ni{} and \Co{}. For this excitation channel to be effective, plumes of \Ni{}-rich material must have been able to reach the outer He-rich layers of the progenitor star of \sn{}, consistent with the results from our two-zone modeling of the bolometric emission from \sn{} in \S \ref{sec:Lbol}, and consistent with the results from recent 3D simulations of stellar explosions (e.g., \citealt{Wongwathanarat2015}) and studies of SN remnants \citep{Milisavljevic15b}. Indeed, models of \cite{Dessart2012} showed that a single plume of \Ni{}-rich material injected into the outer layers (e.g., in the form of a jet) is capable of producing weak He features. Alternatively, 3D simulations of neutrino-driven explosions have shown mixing instabilities that are capable of injecting \Ni{}- and Si-rich plumes to the higher velocities layers of the ejecta \citep[e.g.,][]{Hammer2010}. Indeed, we observed a very similar velocity evolution for \ion{Si}{ii} and \ion{He}{i}, supporting this scenario. In their spectral modelling, \cite{Prentice17} also showed that at early phases heavy ions were travelling at a very high speed; \ion{Fe}{ii} was the fastest species at 26~000 km s$^{-1}$. Also the \ion{Ca}{ii} showed similar high velocities, possibly hinting to some high velocity material coming from the progenitor core, as seen in GRB-SNe \citep[e.g.][]{Bufano12,Toy2016,Ashall2019}.

Among type Ic SNe with broad lines, SNe 2009bb and 2012ap have been reported to have signatures of He in their spectra\footnote{There is also a disputed claim of He in SN 1998bw \citep{Patat2001}, and in the more recent SN 2017iuk \citep{Izzo2019}.} \citep{Pignata2011,Milisavljevic15}. Interestingly, SNe 2009bb and 2012ap are currently the only two cases of SNe with mildly relativistic ejecta not associated with a GRB \citep{Soderberg10,Margutti14b,Chakraborti15}. This phenomenology has been suggested to be the result of a jet-driven stellar explosion where the jet fails to break through the stellar envelope \citep{Morsony07,Lazzati12,Margutti14b}. In this picture, relativistic SNe and GRBs are intrinsically different types of explosions, as opposed to similar explosions viewed from different perspectives. In relativistic SNe the jet is possibly choked by the more extended stellar envelope, but manages to ``transport'' some \Ni{}-rich material outwards and then excites some residual He that the stellar progenitor failed to shed before stellar death \citep{Maeda2002,Suzuki2018,Izzo2019}.

We speculate that a similar scenario applies to \sn{}, for which the lack of evidence for mildly relativistic ejecta can be explained as the result of a jet that died deep inside the star, leaving no imprint on the dynamics of the fastest ejected material, yet accelerating the inner layers to velocities larger than in normal type Ic SNe (Fig. \ref{fig:modjaz}), and at the same time injecting metal-rich material into the outer layer of the ejecta. In this context, the difference between normal type Ibc SNe and those with large ejecta velocities (including type Ic-BL and \sn{}) would be ascribed to the absence/presence of a jet at the time of core-collapse \citep{Khokhlov1999,Granot04,Wheeler10,Lazzati12,Nagakura12,Margutti14b,Soker2016}.

\section{Summary and Conclusions} \label{sec:concl}
We present the results of a multi-wavelength, $\gamma$-rays to radio campaign on the peculiar \sn{} \citep{Yamanaka17,Kumar18,Prentice17} during its first 420 days of evolution. Our findings can be summarized as follows:
\begin{itemize}
 \item From extensive UV/optical/NIR photometry we derive a broad bolometric light-curve (Fig. \ref{Fig:LCcomparison}), which is suggestive of a larger-than-average explosion ejecta mass. From our two-zone modeling we infer $M_{ej,tot}\sim 4-7\,\rm{M_{\sun}}$ (consistent with previous findings by \citealt{Kumar18}), with a larger fraction of \Ni{} per unit mass in the outer part of the ejecta. We also constrain a total kinetic energy of $E_{k}\sim (7-8)\times 10^{51}\,\rm{erg}$.
 \item Our spectroscopic analysis supports the presence of He in the SN ejecta, confirming the previous findings by \cite{Yamanaka17}, \cite{Kumar18} and \cite{Prentice17}. We furthermore find a low [\ion{Ca}{ii}] \lam\lam7291,7324 to [\ion{O}{i}] \lam\lam6300,6364 ratio, suggestive of a large progenitor core mass at the time of collapse.
 \item \sn{} is a luminous source of radio and X-rays, which result from the propagation of a sub-relativistic blast wave with $v\sim 0.15c$ into a dense environment sculpted by sustained mass-loss from the progenitor star before core-collapse. We infer a lower limit on the mass-loss rate of $\dot M \sim (3-4)\times 10^{-5}\,\rm{M_{\odot}yr^{-1}}$ (for wind velocity $v_w=1000\,\rm{km\,s^{-1}}$ and assuming energy equipartion), significantly larger than in type Ic-BL SNe. 
 \item Radio modelling also revealed a phase of higher mass-loss rate lasting until $\sim30$~years before explosion. Additionally, we inferred a post-explosion density profile of the outer ejecta compatible with the explosion of a compact star (e.g., a WR, as opposed to extended progenitors like red and yellow supergiant stars).
 \item We investigated the presence of a high-energy prompt pulse of emission in the $\gamma$-rays. From our analysis we can rule out a SN-associated shock breakout pulse with energy $E_{\gamma}>2\times 10^{46}\,\rm{erg}$, consistent with the theoretical expectations of shock break out from WR stars or from extended winds.
\end{itemize}

The emerging picture is that of a massive compact progenitor star that was able to retain some He until collapse, despite the heavy mass loss experienced in the years leading up to stellar demise. The combination of (i) large ejecta mass and (ii) large mass-loss in a H-stripped core-collapse SN with (iii) weak He features in the spectra, set \sn{} apart from all SNe with similar data coverage and quality in the literature. We speculate that the energetic \sn{} might be the result of a failed jet that was choked by the extended envelope mass of its progenitor star, in analogy with the relativistic type Ic-BL SNe 2009bb and 2012ap for which He has been identified in the ejecta. It is possible that this picture of a jet-driven explosion where the jet has been choked while trying to pierce through the He-rich stellar envelope extends to the entire class of type Ic-BL SNe that are not associated with GRBs. Future observing campaigns of type Ic-BL SNe with coordinated optical and NIR spectroscopy will reveal if traces of He in type Ic-BL SNe are more common than currently thought.

\section*{Acknowledgments}
We thank N. Morrell for observing at du Pont telescope, and E. Falco for observing at FLWO.

ASAS-SN is supported by the Gordon and Betty Moore Foundation through grant GBMF5490 to the Ohio State University and NSF grant AST-1515927. Development of ASAS-SN has been supported by NSF grant AST-0908816, the Mt. Cuba Astronomical Foundation, the Center for Cosmology and AstroParticle Physics at the Ohio State University, the Chinese Academy of Sciences South America Center for Astronomy (CASSACA), the Villum Foundation, and George Skestos. We thank the Las Cumbres Observatory (LCOGT) and its staff for its continuing support of the ASAS-SN project.
The Liverpool Telescope is operated on the island of La Palma by Liverpool John Moores University in the Spanish Observatorio del Roque de los Muchachos of the Instituto de Astrof\'{\i}sica de Canarias with financial support from the UK Science and Technology Facilities Council.
Based on observations made with the Nordic Optical Telescope Scientific Association at the Observatorio Roque de los Muchachos, La Palma, Spain, of the Instituto de Astrof\'{\i}sica de Canarias.
Partially based on observations obtained with Copernico 1.82m Telescope (Asiago) operated by INAF Osservatorio Astronomico di Padova.
This work makes use of data gathered with the 6.5-m Magellan Telescopes, at Las Campanas Observatory in Chile.
Based on observations obtained with XMM-Newton, an ESA science mission with instruments and contributions directly funded by ESA Member States and NASA.
We acknowledge the use of public data from the Swift data archive.
This work made use of the data products generated by the NYU SN group, and 
released under DOI:10.5281/zenodo.58766, 
available at \url{https://github.com/nyusngroup/SESNtemple/}.
This work has made use of the Weizmann Interactive Supernova Data Repository (\url{https://wiserep.weizmann.ac.il}).
This work has made use of the Berkeley Supernova Database (\url{http://heracles.astro.berkeley.edu/sndb/}).
This research has made use of the NASA/IPAC Extragalactic Database (NED), which is operated by the Jet Propulsion Laboratory, California Institute of Technology, under contract with NASA.
The National Radio Astronomy Observatory is a facility of the National Science Foundation operated under cooperative agreement by Associated Universities, Inc.
The RM group at Northwestern is partially supported by the National Aeronautics and Space Administration through Chandra Award Number GO6-17053A issued by the Chandra X-ray Center, which is operated by the Smithsonian Astrophysical Observatory for and on behalf of the National Aeronautics Space Administration under contract NAS8-03060, the Swift Guest Investigator program, NASA Grants NNX17AD82G and 80NSSC18K0575 and the XMM Guest Investigator program, NASA Grant NNX16AT58G.
SD and PC acknowledge Project 11573003 supported by NSFC. This research uses data obtained through the Telescope Access Program (TAP), which has been funded by the National Astronomical Observatories of China, the Chinese Academy of Sciences, and the Special Fund for Astronomy from the Ministry of Finance.
KH is grateful for support under the Fermi Guest Investigator program, NASA Grant NNX15AU74G.
Support for JLP is provided in part by FONDECYT through the grant 1191038 and by the Ministry of Economy, Development, and Tourism’s Millennium Science Initiative through grant IC120009, awarded to The Millennium Institute of Astrophysics, MAS.
LT and SB are partially supported by PRIN INAF 2017 ``Towards the SKA and CTA era: discovery, localisation and physics of transient sources (PI M. Giroletti).''
MS is supported by a generous grant (13261) from VILLUM FONDEN and a project grant (8021-00170B) from the Independent Research Fund Denmark. 
NER acknowledges support from the Spanish MICINN grant ESP2017-82674-R and FEDER funds.
CG is supported by a VILLUM FONDEN Investigator grant (project number 16599).
JH acknowledges financial support from the Finnish Cultural Foundation.

\facility{ASAS-SN, Liverpool Telescope, Neil Gehrels \emph{Swift} Observatory, XMM-Newton, Nordic Optical Telescope, Multiple Mirror Telescope, Rapid Eye Mount, 1.82m Copernico, 1.22m Galileo, du Pont Telescope, Large Binocular Telescope, Magellan, Telescopio Nazionale Galileo, Tillinghast 1.5m, Las Cumbres Observatory, DEMONEXT, WHO 1m (Weihai Observatory), T50 (OAUV), Gem (University of Iowa)}.


\bibliographystyle{apj}
\bibliography{terreran}

\clearpage
\appendix

\renewcommand\thetable{A\arabic{table}} 
\setcounter{table}{0}

\setlength{\tabcolsep}{0pt}
\begin{table*}[h]
 \centering
 \caption{Telescopes and instruments used for the photometric follow-up of \sn{}. The acronyms reported in the first column are those used in Tables \ref{fig: ugriz}$-$\ref{fig: SDA}.}
\begin{tabular}{c<{\hspace{5pt}}lccr<{\hspace{-\tabcolsep}}>{\hspace{-\tabcolsep}\,}c<{\hspace{-\tabcolsep}\,}>{\hspace{-\tabcolsep}}lc}
 \hline
 Code & Telescope & Instrument & Pixel size & \multicolumn{3}{c}{FoV} & Filters\\
 \hline
ASA & 14cm Brutus\footnote{Operated by the ASAS-SN team \citep{Shappee2014}.} & Fairchild CCD3041 & 7.8\arcsec & 4.47\degr &$\times$& 4.47\degr & \textit{V}\\

\rowcolor{gray!20}
RK & 25cm Meade SCT \footnote{Operated by R. A. Koff at Antelope Hills Observatory, Bennett, CO, USA.} & Apogee AP-47 & 1.02\arcsec & 17.4\arcmin &$\times$& 17.4\arcmin & \textit{BVRI}\\

JB & 43cm PlaneWave CDK \footnote{Operated by Joseph Brimacombe at New Mexico Skies, New Mexico, USA.} &SBIG STL-6303 & 0.63\arcsec & 34\arcmin &$\times$& 22\arcmin & \textit{BVri}\\

\rowcolor{gray!20}
GS & 43cm PlaneWave CDK \footnote{Operated by Geoffrey Stone at Sierra Remote Observatories, Auberry, CA, USA.}& SBIG STXL-11002 & 0.63\arcsec & 42\arcmin &$\times$& 28\arcmin & \textit{BV}\\

T50 &T50 \footnote{Operated by the Astronomical Observatory of the University of Valencia (OAUV) at Aras de los Olmos, Valencia, Spain.}& ProLine PL16801 & 0.54\arcsec & 36.9\arcmin &$\times$&36.9\arcmin & \textit{BVri}\\

\rowcolor{gray!20}
Gem & Gem\footnote{Operated by the University of Iowa at Iowa Robotic Observatory.} & SBIG 6303e & 1.08\arcsec & 27.6\arcmin &$\times$& 18.4\arcmin & \textit{BVgri}\\

DEM & 51cm PlaneWave CDK\footnote{Operated by DEMONEXT \citep{Villanueva2018}.} & Fairchild CCD3041 & 0.90\arcsec & 30.7\arcmin &$\times$& 30.7\arcmin & \textit{BVri}\\
\hline
\end{tabular}
\label{tab: instr}
\end{table*}

\setlength{\tabcolsep}{5pt}

\renewcommand{\arraystretch}{0.9}

\begin{table*}
\centering
\caption{\textit{ugriz} photometry.}
\begin{tabular}{>{\centering}m{1.2cm}>{\centering}m{1.8cm}>{\centering}m{1.8cm}>{\centering}m{1.8cm}>{\centering}m{1.8cm}>{\centering}m{1.8cm}>{\centering}m{3cm}}
\hline 
MJD & \textit{u} & \textit{g} & \textit{r} & \textit{i} & \textit{z} & Instrument\tabularnewline 
 & (mag) & (mag) & (mag) & (mag) & (mag) & \tabularnewline 
\hline 
57536.14 & $-$ & $-$ & 15.20 (0.04) & 15.46 (0.03) & $-$ & NOT\tabularnewline 
57536.40 & $-$ & 15.80 (0.05) & 15.17 (0.05) & 15.55 (0.10) & $-$ & Gem\tabularnewline 
57537.15 & $-$ & $-$ & 14.84 (0.15) & 15.09 (0.11) & $-$ & NOT\tabularnewline 
57537.41 & $-$ & 15.43 (0.04) & 14.87 (0.05) & 15.25 (0.10) & $-$ & Gem\tabularnewline 
57538.39 & $-$ & 15.17 (0.04) & 14.63 (0.05) & 15.03 (0.10) & $-$ & Gem\tabularnewline 
57539.39 & $-$ & 14.93 (0.04) & 14.43 (0.05) & 14.83 (0.10) & $-$ & Gem\tabularnewline
\end{tabular}
\end{table*}

\addtocounter{table}{-1}

\begin{table*}
\centering
\caption{\textit{Continued:} \textit{ugriz} photometry.}
\begin{tabular}{>{\centering}m{1.2cm}>{\centering}m{1.8cm}>{\centering}m{1.8cm}>{\centering}m{1.8cm}>{\centering}m{1.8cm}>{\centering}m{1.8cm}>{\centering}m{3cm}}
\hline 
MJD & \textit{u} & \textit{g} & \textit{r} & \textit{i} & \textit{z} & Instrument\tabularnewline 
 & (mag) & (mag) & (mag) & (mag) & (mag) & \tabularnewline 
\hline 
57558.40 & $-$ & $-$ & 13.68 (0.06) & 13.94 (0.11) & $-$ & LCOGT\tabularnewline 
57559.07 & 16.89 (0.05) & 14.60 (0.03) & 13.65 (0.03) & 13.95 (0.03) & 13.59 (0.03) & LT\tabularnewline 
57559.24 & $-$ & $-$ & 13.70 (0.08) & 13.90 (0.13) & $-$ & REM\tabularnewline 
57559.44 & $-$ & $-$ & 13.73 (0.03) & 13.97 (0.05) & $-$ & LCOGT\tabularnewline 
57560.30 & $-$ & 14.76 (0.35) & 13.74 (0.08) & 13.94 (0.11) & $-$ & REM\tabularnewline 
57560.40 & $-$ & $-$ & $-$ & 14.00 (0.05) & $-$ & LCOGT\tabularnewline
\hline 
\end{tabular}
\end{table*}

\addtocounter{table}{-1}

\begin{table*}
\centering
\caption{\textit{Continued:} \textit{ugriz} photometry.}
\begin{tabular}{>{\centering}m{1.2cm}>{\centering}m{1.8cm}>{\centering}m{1.8cm}>{\centering}m{1.8cm}>{\centering}m{1.8cm}>{\centering}m{1.8cm}>{\centering}m{3cm}}
\hline 
MJD & \textit{u} & \textit{g} & \textit{r} & \textit{i} & \textit{z} & Instrument\tabularnewline 
 & (mag) & (mag) & (mag) & (mag) & (mag) & \tabularnewline 
\hline 
57597.07 & $-$ & $-$ & 15.16 (0.28) & $-$ & $-$ &T50\tabularnewline 
57597.73 & $-$ & $-$ & 15.18 (0.04) & 15.07 (0.06) & $-$ & LCOGT\tabularnewline 
57598.20 & 18.80 (0.04) & 16.29 (0.03) & 15.13 (0.04) & 15.08 (0.04) & 14.46 (0.03) & LT\tabularnewline 
57598.72 & $-$ & $-$ & 15.20 (0.04) & 15.08 (0.06) & $-$ & LCOGT\tabularnewline 
57599.70 & $-$ & $-$ & 15.24 (0.04) & 15.09 (0.06) & $-$ & LCOGT\tabularnewline 
57601.14 & 18.92 (0.07) & 16.40 (0.03) & 15.25 (0.04) & 15.14 (0.03) & $-$ & NOT\tabularnewline
\hline 
\end{tabular}
\end{table*}

\begin{table*}
\centering
\caption{\textit{UBVRI} photometry.}
\begin{tabular}{>{\centering}m{1.2cm}>{\centering}m{1.8cm}>{\centering}m{1.8cm}>{\centering}m{1.8cm}>{\centering}m{1.8cm}>{\centering}m{1.8cm}>{\centering}m{3cm}}
\hline 
MJD & \textit{U} & \textit{B} & \textit{V} & \textit{R} & \textit{I} & Instrument\tabularnewline 
 & (mag) & (mag) & (mag) & (mag) & (mag) & \tabularnewline 
\hline 
57527.53 & $-$ & $-$ & $>$17.43 & $-$ & $-$ & ASA\tabularnewline 
57529.54 & $-$ & $-$ & $>$17.29 & $-$ & $-$ & ASA\tabularnewline 
57535.55 & $-$ & $-$ & 15.74 (0.07) & $-$ & $-$ & ASA\tabularnewline 
57535.77 & 15.86 (0.06) & 16.51 (0.15) & 15.86 (0.09) & $-$ & $-$ & UVOT\tabularnewline 
57536.14 & $-$ & 16.30 (0.03) & 15.51 (0.03) & $-$ & $-$ & NOT\tabularnewline 
57536.36 & $-$ & 16.29 (0.04) & 15.46 (0.03) & 15.10 (0.04) & 15.19 (0.07) & RK\tabularnewline
\end{tabular}
\end{table*}

\addtocounter{table}{-1}

\begin{table*}
\centering
\caption{\textit{Continued:} \textit{UBVRI} photometry.}
\begin{tabular}{>{\centering}m{1.2cm}>{\centering}m{1.8cm}>{\centering}m{1.8cm}>{\centering}m{1.8cm}>{\centering}m{1.8cm}>{\centering}m{1.8cm}>{\centering}m{3cm}}
\hline 
MJD & \textit{U} & \textit{B} & \textit{V} & \textit{R} & \textit{I} & Instrument\tabularnewline 
 & (mag) & (mag) & (mag) & (mag) & (mag) & \tabularnewline 
\hline 
57553.37 & $-$ & 14.99 (0.04) & 13.74 (0.04) & $-$ & $-$ & DEM\tabularnewline 
57554.20 & 15.45 (0.06) & 15.05 (0.06) & 13.73 (0.05) & $-$ & $-$ & UVOT\tabularnewline 
57554.30 & $-$ & 15.09 (0.05) & 13.77 (0.04) & 13.47 (0.05) & 13.68 (0.08) & RK\tabularnewline 
57554.42 & $-$ & 15.05 (0.04) & 13.76 (0.04) & $-$ & $-$ & DEM\tabularnewline 
57555.34 & $-$ & 15.12 (0.04) & 13.82 (0.03) & 13.46 (0.05) & 13.63 (0.07) & RK\tabularnewline 
57555.56 & 15.55 (0.06) & 15.07 (0.06) & 13.78 (0.04) & $-$ & $-$ & UVOT\tabularnewline
\hline 
\end{tabular}
\end{table*}

\addtocounter{table}{-1}

\begin{table*}
\centering
\caption{\textit{Continued:} \textit{UBVRI} photometry.}
\begin{tabular}{>{\centering}m{1.2cm}>{\centering}m{1.8cm}>{\centering}m{1.8cm}>{\centering}m{1.8cm}>{\centering}m{1.8cm}>{\centering}m{1.8cm}>{\centering}m{3cm}}
\hline 
MJD & \textit{U} & \textit{B} & \textit{V} & \textit{R} & \textit{I} & Instrument\tabularnewline 
 & (mag) & (mag) & (mag) & (mag) & (mag) & \tabularnewline 
\hline 
57581.30 & 17.59 (0.14) & 16.74 (0.14) & 15.20 (0.06) & $-$ & $-$ & UVOT\tabularnewline 
57582.13 & $-$ & 16.85 (0.07) & 15.24 (0.04) & $-$ & $-$ & LCOGT\tabularnewline 
57582.34 & $-$ & 16.77 (0.10) & 15.22 (0.13) & 14.49 (0.13) & 14.28 (0.17) & JB\tabularnewline 
57583.13 & $-$ & $-$ & 15.28 (0.04) & $-$ & $-$ & LCOGT\tabularnewline 
57583.27 & $-$ & 16.66 (0.06) & 15.24 (0.06) & 14.47 (0.25) & 14.25 (0.15) & RK\tabularnewline 
57583.42 & $-$ & $-$ & 15.38 (0.04) & $-$ & $-$ & ASA\tabularnewline
\hline 
\end{tabular}
\end{table*}

\addtocounter{table}{-1}

\begin{table*}
\centering
\caption{\textit{Continued:} \textit{UBVRI} photometry.}
\begin{tabular}{>{\centering}m{1.2cm}>{\centering}m{1.8cm}>{\centering}m{1.8cm}>{\centering}m{1.8cm}>{\centering}m{1.8cm}>{\centering}m{1.8cm}>{\centering}m{3cm}}
\hline 
MJD & \textit{U} & \textit{B} & \textit{V} & \textit{R} & \textit{I} & Instrument\tabularnewline 
 & (mag) & (mag) & (mag) & (mag) & (mag) & \tabularnewline 
\hline 
57611.53 & $-$ & $-$ & 15.97 (0.05) & $-$ & $-$ & ASA\tabularnewline 
57612.40 & $-$ & 17.16 (0.05) & 15.89 (0.01) & $-$ & $-$ & GS\tabularnewline 
57612.65 & $-$ & $-$ & 15.91 (0.04) & $-$ & $-$ & LCOGT\tabularnewline 
57613.20 & $-$ & 17.17 (0.06) & 15.78 (0.11) & 15.24 (0.24) & 14.90 (0.18) & RK\tabularnewline 
57613.31 & $-$ & $-$ & 15.90 (0.06) & $-$ & $-$ & ASA\tabularnewline 
57613.51 & 17.96 (0.21) & 17.26 (0.20) & 15.94 (0.10) & $-$ & $-$ & UVOT\tabularnewline
\hline 
\end{tabular}
\end{table*}

\addtocounter{table}{-1}

\begin{table*}
\centering
\caption{\textit{Continued:} \textit{UBVRI} photometry.}
\begin{tabular}{>{\centering}m{1.2cm}>{\centering}m{1.8cm}>{\centering}m{1.8cm}>{\centering}m{1.8cm}>{\centering}m{1.8cm}>{\centering}m{1.8cm}>{\centering}m{3cm}}
\hline 
MJD & \textit{U} & \textit{B} & \textit{V} & \textit{R} & \textit{I} & Instrument\tabularnewline 
 & (mag) & (mag) & (mag) & (mag) & (mag) & \tabularnewline 
\hline 
57662.15 & $-$ & 17.73 (0.08) & 16.70 (0.03) & $-$ & $-$ & NOT\tabularnewline 
57662.69 & 18.81 (0.31) & 17.84 (0.33) & 16.93 (0.15) & $-$ & $-$ & UVOT\tabularnewline 
57663.90 & $-$ & $-$ & 16.75 (0.06) & $-$ & $-$ & LCOGT\tabularnewline 
57665.84 & $-$ & 17.96 (0.09) & 16.77 (0.04) & $-$ & $-$ & LCOGT\tabularnewline 
57667.42 & $-$ & $-$ & 16.80 (0.11) & $-$ & $-$ & ASA\tabularnewline 
57667.86 & $-$ & 17.85 (0.10) & 16.75 (0.05) & $-$ & $-$ & LCOGT\tabularnewline
\hline 
\end{tabular}
\end{table*}

\begin{table*}
\centering
\caption{\textit{JHK} photometry.}
\begin{tabular}{>{\centering}m{1.2cm}>{\centering}m{1.8cm}>{\centering}m{1.8cm}>{\centering}m{1.8cm}>{\centering}m{3cm}}
\hline 
MJD & \textit{J} & \textit{H} & \textit{K} & Instrument\tabularnewline 
 & (mag) & (mag) & (mag) & \tabularnewline 
\hline 
57540.17 & $-$ & 13.78 (0.27) & $-$ & LT\tabularnewline 
57542.21 & $-$ & 13.55 (0.30) & $-$ & LT\tabularnewline 
57546.29 & 13.33 (0.03) & 13.21 (0.02) & 13.14 (0.03) & REM\tabularnewline 
57547.37 & 13.30 (0.02) & 13.12 (0.02) & 13.00 (0.04) & REM\tabularnewline 
57548.38 & 13.26 (0.04) & 13.08 (0.02) & 12.94 (0.04) & REM\tabularnewline 
57549.42 & 13.25 (0.04) & 13.09 (0.02) & 12.98 (0.06) & REM\tabularnewline
\hline 
\end{tabular}
\label{fig: JHK}
\end{table*}

\addtocounter{table}{-1}

\begin{table*}
\centering
\caption{\textit{Continued:} \textit{JHK} photometry.}
\begin{tabular}{>{\centering}m{1.2cm}>{\centering}m{1.8cm}>{\centering}m{1.8cm}>{\centering}m{1.8cm}>{\centering}m{3cm}}
\hline 
MJD & \textit{J} & \textit{H} & \textit{K} & Instrument\tabularnewline 
 & (mag) & (mag) & (mag) & \tabularnewline 
\hline 
57623.15 & 15.11 (0.40) & 14.69 (0.34) & 14.19 (0.39) & NOT\tabularnewline 
57629.11 & 15.47 (0.06) & 14.72 (0.07) & 14.70 (0.17) & REM\tabularnewline 
57634.96 & $-$ & 14.80 (0.28) & $-$ & LT\tabularnewline 
57636.95 & $-$ & 14.82 (0.30) & $-$ & LT\tabularnewline 
57638.94 & $-$ & 14.89 (0.28) & $-$ & LT\tabularnewline 
57642.93 & $-$ & 15.02 (0.28) & $-$ & LT\tabularnewline 
57650.92 & $-$ & 15.17 (0.29) & $-$ & LT\tabularnewline 
57654.96 & $-$ & 15.32 (0.25) & $-$ & LT\tabularnewline 
57657.92 & $-$ & 15.42 (0.23) & $-$ & LT\tabularnewline 
57663.93 & $-$ & 15.58 (0.26) & $-$ & LT\tabularnewline 
57666.87 & $-$ & 15.58 (0.22) & $-$ & LT\tabularnewline 
57669.88 & $-$ & 15.64 (0.31) & $-$ & LT\tabularnewline 
57672.86 & $-$ & 15.69 (0.30) & $-$ & LT\tabularnewline 
57675.84 & $-$ & 15.70 (0.28) & $-$ & LT\tabularnewline 
57678.87 & $-$ & 15.87 (0.30) & $-$ & LT\tabularnewline 
57681.88 & $-$ & 15.87 (0.28) & $-$ & LT\tabularnewline 
57689.87 & $-$ & 15.97 (0.27) & $-$ & LT\tabularnewline 
57693.83 & $-$ & 16.02 (0.23) & $-$ & LT\tabularnewline 
57700.84 & $-$ & 16.15 (0.26) & $-$ & LT\tabularnewline 
57708.86 & $-$ & 16.38 (0.26) & $-$ & LT\tabularnewline 
57711.85 & $-$ & 16.33 (0.27) & $-$ & LT\tabularnewline 
57732.87 & $-$ & 16.71 (0.26) & $-$ & LT\tabularnewline 
57856.24 & $-$ & 18.56 (0.35) & $-$ & LT\tabularnewline 
57863.21 & $-$ & $>$17.65 & $-$ & LT\tabularnewline 
57866.22 & $-$ & 18.61 (0.35) & $-$ & LT\tabularnewline 
\hline 
\end{tabular}
\end{table*}

\begin{table*}
\centering
\renewcommand{\thefootnote}{\fnsymbol{footnote}}\caption{UV photometry.}
\begin{tabular}{>{\centering}m{1.2cm}>{\centering}m{1.8cm}>{\centering}m{1.8cm}>{\centering}m{1.8cm}>{\centering}m{3cm}}
\hline 
MJD & \textit{UV-W2} & \textit{UV-M2} & \textit{UV-W1} & Instrument\tabularnewline 
 & (mag) & (mag) & (mag) & \tabularnewline 
\hline 
57535.77 & 17.74 (0.26) & 17.71 (0.22) & 16.74 (0.17) & UVOT\tabularnewline 
57539.09 & 17.11 (0.21) & 17.06 (0.14) & 16.07 (0.13) & UVOT\tabularnewline 
57542.61 & 17.01 (0.16) & 17.26 (0.16) & 15.88 (0.10) & UVOT\tabularnewline 
57543.98 & 17.02 (0.16) & 17.50 (0.18) & 15.98 (0.11) & UVOT\tabularnewline 
57545.94 & 17.25 (0.19) & 17.58 (0.20) & 16.08 (0.11) & UVOT\tabularnewline 
57547.47 & 17.31 (0.19) & 17.84 (0.23) & 16.28 (0.12) & UVOT\tabularnewline
\end{tabular}
\end{table*}

\begin{table*}
 \centering
\caption{Peak time and peak magnitudes of \sn{} in all filters. The errors reported are the statistical error on the identification of the maxima, and do not include the uncertainties on the discovery (MJD 57535.55) and explosion date (MJD 57531.9).}
\begin{tabular}{ccccc}
\hline
Band & MJD max & \multicolumn{2}{c}{Phase max [d]} & Mag max \\
&& (from disc.) & (from expl.) &\\ 
\hline
\textit{UV-W2} & $57541.50\pm 4.13$ & ~$5.95 $ & ~$9.60 $ & $16.99 \pm 0.17$\\
\textit{UV-M2} & $57540.13\pm 4.34$ & ~$4.58 $ & ~$8.23 $ & $17.01 \pm 0.98$\\
\textit{UV-W1} & $57542.01\pm 0.95$ & ~$6.46 $ & $10.11 $ & $15.90 \pm 0.07$\\
\textit{u} & 	 $57543.51\pm 0.52$ & ~$7.96 $ & $11.61 $ & $15.61 \pm 0.02$\\
\textit{U} & 	 $57543.85\pm 2.21$ & ~$8.30 $ & $11.95 $ & $14.68 \pm 0.06$\\
\textit{B} & 	 $57547.87\pm 0.50$ & $12.32 $ & $15.97 $ & $14.83 \pm 0.02$\\
\textit{g} & 	 $57548.85\pm 2.83$ & $13.30 $ & $16.95 $ & $14.15 \pm 0.09$\\
\textit{V} & 	 $57550.24\pm 0.16$ & $14.69 $ & $18.34 $ & $17.86 \pm 0.01$\\
\textit{r} & 	 $57551.88\pm 0.36$ & $16.33 $ & $19.98 $ & $13.59 \pm 0.01$\\
\textit{R} & 	 $57551.10\pm 0.60$ & $15.55 $ & $19.20 $ & $13.41 \pm 0.04$\\
\textit{i} & 	 $57554.38\pm 0.86$ & $18.83 $ & $22.48 $ & $13.93 \pm 0.02$\\
\textit{I} & 	 $57553.54\pm 2.26$ & $17.99 $ & $21.64 $ & $13.63 \pm 0.06$\\
\textit{z} & 	 $57555.50\pm 0.67$ & $19.95 $ & $23.60 $ & $13.58 \pm 0.01$\\
\textit{J} & 	 $57558.74\pm 0.84$ & $23.19 $ & $26.84 $ & $13.03 \pm 0.03$\\
\textit{H} & 	 $57562.30\pm 2.75$ & $26.75 $ & $30.40 $ & $12.85 \pm 0.09$\\
\textit{K} & 	 $57560.88\pm 2.13$ & $25.33 $ & $28.98 $ & $12.72 \pm 0.03$\\
\hline
\end{tabular}
\label{tab: max}
\end{table*}

\setlength{\tabcolsep}{0pt}
\begin{table*}
 \centering
 \caption{Telescopes, instruments and configurations used for the spectroscopic follow-up of \sn{}.}
 \begin{tabular}{lccccr<{\hspace{-\tabcolsep}}>{\hspace{-\tabcolsep}\,}c<{\hspace{-\tabcolsep}\,}>{\hspace{-\tabcolsep}}l}
 \hline
 Telescope & Instrument & Grism/Grating & Slit &Resolution [R]& \multicolumn{3}{c}{Wavelength range}\\
 \hline
\rowcolor{gray!20} 1.22m Galileo & B\&C spectrograph & 300 ln/mm&3.93\arcsec&636&3800&$-$&8000~\AA\\
 & & Gr\#4&&363&3360&$-$&7740~\AA\\
 & & VPH6&&300&4500&$-$&10000~\AA\\
 \multirow{-3}{*}{1.82m Copernico} & \multirow{-3}{*}{AFOSC}& VPH7&\multirow{-3}{*}{1.69\arcsec}&375&3200&$-$&7300~\AA\\
\rowcolor{gray!20} du Pont Telescope&B\&C spectrograph&300 ln/mm&2.71\arcsec&1667&3300&$-$&9500~\AA\\
 &&Red&&2300&5500&$-$&10500~\AA\\
 \multirow{-2}{*}{Large Binocular Telescope (LBT)}&\multirow{-2}{*}{MODS}&Blue&\multirow{-2}{*}{0.6\arcsec}&1850&3200&$-$&6000~\AA\\
\rowcolor{gray!20} Magellan & IMACS &300 ln/mm& 0.9\arcsec &1100&3650&$-$&9740~\AA\\
 & &VPH-Blue &1.2\arcsec&1600&3900&$-$&6800~\AA\\
 & & &1.0\arcsec&1600&3920&$-$&9050~\AA\\
 \multirow{-3}{*}{MDM Hiltner 2.4 m} & \multirow{-3}{*}{OSMOS}&\multirow{-2}{*}{VPH-Red}&1.2\arcsec&1500&5350&$-$&10500~\AA\\
\rowcolor{gray!20} Multiple Mirror Telescope (MMT)&Blue Channel&300 ln/mm&1.0\arcsec&750&3300&$-$&8600~\AA\\
LT& SPRAT & Blue & 1.8\arcsec &350&4000 &$-$& 8000~\AA\\
\rowcolor{gray!20}NOT &ALFOSC& Gr\#4& 1.0\arcsec& 360& 3200&$-$&9600~\AA\\
Tillinghast 1.5m (FLWO)& FAST & 300 ln/mm & 3.0\arcsec & 900 & 3530&$-$&7470~\AA\\
\rowcolor{gray!20} & & LR-B & & 585& 3000&$-$&8430~\AA\\
\rowcolor{gray!20}\multirow{-2}{*}{TNG}&\multirow{-2}{*}{DOLORES}&LR-R&\multirow{-2}{*}{1.0\arcsec}&714&4470&$-$&10070~\AA\\

\hline
\end{tabular}
\label{tab: instr2}
\end{table*}
\setlength{\tabcolsep}{5pt}

\begin{table*}
\centering
\caption{Spectroscopic log.}
\begin{tabular}{ccccc}
\hline 
Date obs. & MJD & Tel.+Inst. & Slit & Grism/Grating\\ 
\hline 
2016$-$05$-$28 & 57536.2 & NOT+ALFOSC & 1.0\arcsec & Gr\#4\\
2016$-$05$-$28 & 57536.2 & LT+SPRAT & 1.8\arcsec & Blue\\
2016$-$05$-$29 & 57537.2 & LT+SPRAT & 1.8\arcsec & Blue\\
2016$-$05$-$29 & 57537.1 & NOT+ALFOSC & 1.0\arcsec & Gr\#4\\
2016$-$05$-$31 & 57539.4 & MDM+OSMOS & 1.2\arcsec & VPH-Blue\\
2016$-$06$-$01 & 57540.1 & LT+SPRAT & 1.8\arcsec & Blue\\
2016$-$06$-$02 & 57541.4 & Till+FAST & 3.0\arcsec & 300 ln/mm\\
\end{tabular}
\end{table*}

\addtocounter{table}{-1}

\begin{table*}
\centering
\caption{\textit{Continued:} Spectroscopic log.}
\begin{tabular}{ccccc}
\hline 
Date obs. & MJD & Tel.+Inst. & Slit & Grism/Grating\\ 
\hline 
2016$-$09$-$24 & 57655.1 & NOT+ALFOSC & 1.0\arcsec & Gr\#4\\
2016$-$09$-$30 & 57662.0 & LT+SPRAT & 1.8\arcsec & Blue\\
2016$-$10$-$01 & 57662.1 & NOT+ALFOSC & 1.0\arcsec & Gr\#4\\
2016$-$10$-$11 & 57672.9 & LT+SPRAT & 1.8\arcsec & Blue\\
2016$-$10$-$13 & 57674.8 & LT+SPRAT & 1.8\arcsec & Blue\\
2016$-$11$-$02 & 57694.1 & LBT+MODS & 0.6\arcsec& Red+Blue\\
2016$-$11$-$01 & 57693.9 & LT+SPRAT & 1.8\arcsec & Blue\\
2016$-$11$-$15 & 57707.8 & MDM+OSMOS & 1.2\arcsec & VPH-Red\\
2016$-$11$-$18 & 57710.8 & LT+SPRAT & 1.8\arcsec & Blue\\
2016$-$12$-$05 & 57727.7 & 1.82+AFOSC & 1.69\arcsec & VHP6+VPH7\\
2016$-$12$-$06 & 57728.8 & 1.82+AFOSC & 1.69\arcsec & VHP6+VPH7\\
2016$-$12$-$10 & 57732.8 & LT+SPRAT & 1.8\arcsec & Blue\\
2017$-$06$-$17 & 57921.8 & MDM+OSMOS & 1.0\arcsec & VPH-Red\\
2017$-$06$-$28 & 57932.4 & MMT+Blue Channel & 1.0\arcsec & 300 ln/mm\\
2017$-$07$-$20 & 57954.2 & Magellan+IMACS & 0.9\arcsec& 300ln/mm\\
\hline 
\end{tabular}
\end{table*}

\begin{table*}
\centering
\caption{Radio observations of \sn{}.}
\begin{tabular}{cccccc}
\hline 
Start Date & Time since First Light & Frequency & & Flux Density & Project \\
(UT) & (days) & (GHz) & (GHz) & (mJy) & \\
\hline
2016-06-03 & 11 & 5.95 & 2.048 & $1.15\pm0.06$ & 16A-477\\
2016-06-03 & 11 & 9.80 & 2.048 & $3.4\pm0.2$ & 16A-477\\
2016-06-03 & 11 & 14.75 & 2.048 & $8.9\pm0.4$ & 16A-477\\
2016-06-03 & 11 & 21.85 & 2.048 & $21\pm1$ & 16A-477\\
2016-06-13 & 21 & 5.95 & 2.048 & $4.8\pm0.2$ & 16A-477\\
2016-06-13 & 21 & 9.80 & 2.048 & $11.3\pm0.6$ & 16A-477\\
2016-06-13 & 21 & 14.75 & 2.048 & $20\pm1$ & 16A-477\\
2016-06-13 & 21 & 21.85 & 2.048 & $24\pm1$ & 16A-477\\
2016-07-08 & 46 & 3.00 & 2.048 & $3.5\pm0.2$ & 16A-477\\
2016-07-08 & 46 & 5.95 & 2.048 & $12.9\pm0.6$ & 16A-477\\
2016-07-08 & 46 & 9.80 & 2.048 & $20\pm1$ & 16A-477\\
2016-07-08 & 46 & 21.85 & 2.048 & $16.7\pm0.8$ & 16A-477\\
2016-09-07 & 106 & 3.00 & 2.048 & $14.7\pm0.7$ & 16A-477\\
2016-09-07 & 106 & 5.95 & 2.048 & $13.5\pm0.7$ & 16A-477\\
2016-09-07 & 106 & 9.80 & 2.048 & $9.6\pm0.5$ & 16A-477\\
2016-09-07 & 106 & 21.85 & 2.048 & $5.3\pm0.3$ & 16A-477\\
2017-02-17 & 270 & 5.50 & 2.048 & $6.2\pm0.3$ & 17A-167\\
2017-02-17 & 270 & 9.00 & 2.048 & $3.7\pm0.2$ & 17A-167\\
2017-02-25 & 278 & 3.00 & 2.048 & $10.1\pm0.5$ & 17A-167\\
\hline
\label{Tab:radio}
\end{tabular}
\end{table*}

\begin{table*}
\centering
\caption{0.3-10 keV X-ray light-curve of \sn{}.}
\begin{tabular}{cccccc}
\hline 
MJD & Time since Explosion & Time Range & Count-rate & Unabsorbed Flux & Instrument \\
& (days) & (days) & $(c\,s^{-1})$ & $(\rm{erg\,s^{-1}cm^{-2}})$ & \\
\hline
57538.0 & 6.1 &(1.8-10.0)& $(3.44\pm 0.85)\times 10^{-3}$ &$(1.44\pm0.36)\times 10^{-13}$ &XRT\\
57543.4 &11.5 &- & & ($1.44\pm 0.61)\times 10^{-13}$ & XMM\\
57545.6 & 13.7&(10.1-16.8)& $(2.43\pm 0.67)\times 10^{-3}$ & $(1.03\pm0.28)\times 10^{-13} $&XRT\\
57553.3 & 21.4&(17.9-24.0)& $(2.25\pm 0.61 )\times 10^{-3}$ &$(0.95\pm0.25)\times 10^{-13}$&XRT\\
57559.4 & 27.5& -& & $(0.64 \pm 0.19)\times 10^{-13}$ & XMM\\ 
57626.0 & 94.1 &(26.6-157.6)& $(7.47\pm 0.18)\times 10^{-3}$& $(0.31\pm0.08)\times 10^{-13}$&XRT\\
57785.1 & 253.2&(168.9-326.9)& $< 1.54 \times 10^{-3}$ &$<0.65\times 10^{-13} $&XRT\\
\hline
\label{Tab:xrays}
\end{tabular}
\end{table*}

\end{document}